\documentclass[aps,pra,twocolumn,superscriptaddress,floatfix,longbibliography]{revtex4-1}

\usepackage{graphicx}% Include figure files
\usepackage{dcolumn}% Align table columns on decimal point
\usepackage{bm}% bold mathManuscript Title:\\
\usepackage{upgreek}
\usepackage[acronym,nonumberlist]{glossaries} % abbreviations and nomenclature
\usepackage[dvipsnames]{xcolor} % text highlights etc.
\usepackage{soul} % a different highlight package
\usepackage{listings} % for GA code snippet
\usepackage{courier} % for inline code (TurboGAP gamma_p)
\usepackage{url} % to cite the CELD website

\newacronym{DFT}{DFT}{density functional theory}
\newacronym{GGA}{GGA}{generalized gradient approximation}
\newacronym{LDA}{LDA}{local-density approximation}
\newacronym{EAM}{EAM}{embedded atom method}
\newacronym{BOP}{BOP}{bond-order potential}
\newacronym{GAP}{GAP}{Gaussian approximation potential}
\newacronym{SOAP}{SOAP}{smooth  overlap  of  atomic  positions}
\newacronym{MD}{MD}{molecular dynamics}
\newacronym{RMSE}{RMSE}{root mean squared error}
\newacronym{MAE}{MAE}{mean absolute error}
\newacronym{NEB}{NEB}{nudged elastic band}
\newacronym{XRD}{XRD}{X-ray diffraction}

\newacronym{bcc}{bcc}{body-centered cubic}
\newacronym{fcc}{fcc}{face-centered cubic}
\newacronym{hcp}{hcp}{hexagonal close-packed}
\newacronym{sc}{sc}{simple cubic}

\newacronym{FM}{FM}{ferromagnetic}
\newacronym{AFM}{AFM}{antiferromagnetic}
\newacronym{FMLS}{FMLS}{ferromagnetic low spin}
\newacronym{FMHS}{FMHS}{ferromagnetic high spin}
\newacronym{NM}{NM}{nonmagnetic}
\newacronym{AFMD}{AFMD}{antiferromagnetic double layer}

\newacronym{GA}{GA}{genetic algorithm}
\newacronym{ASE}{ASE}{the Atomic Simulation Environment}

\newacronym{PES}{PES}{potential energy surface}
\newacronym{SM}{SM}{Supplemental Material}
\newacronym{CELD}{CELD}{Cambridge Energy Landscape Database}
\newacronym{NP}{NP}{nanoparticle}
\newacronym{ML}{ML}{machine learning}
\newacronym{MDS}{MDS}{multidimensional scaling}
\newacronym{MLP}{MLP}{\gls{ML} potential}

\newacronym{AIMD}{AIMD}{\emph{ab initio} \gls{MD}}
\newacronym{RDF}{RDF}{radial distribution function}

\usepackage{xcolor}

\newcommand{\etal}{\textit{et al.}}
\newcommand{\fig}[1]{Fig.~\ref{#1}}

\begin{document}

\title{Searching for iron nanoparticles with a \\ general-purpose Gaussian approximation potential}

\author{Richard Jana}
\email{richard.jana@aalto.fi}
\affiliation{Department of Chemistry and Materials Science, Aalto University, 02150, Espoo, Finland}

\author{Miguel A. Caro}
\affiliation{Department of Chemistry and Materials Science, Aalto University, 02150, Espoo, Finland}

\date{\today}

\begin{abstract}
We present a general-purpose machine learning Gaussian approximation potential (GAP) for iron that is applicable to all bulk crystal structures found experimentally under diverse thermodynamic conditions, as well as surfaces and nanoparticles (NPs). By studying its phase diagram, we show that our GAP remains stable at extreme conditions, including those found in the Earth's core. The new GAP is particularly accurate for the description of NPs. We use it to identify new low-energy NPs, whose stability is verified by performing density functional theory calculations on the GAP structures. Many of these NPs are lower in energy than those previously available in the literature up to $N_\text{atoms}=100$. We further extend the convex hull of available stable structures to $N_\text{atoms}=200$. For these NPs, we study characteristic surface atomic motifs using data clustering and low-dimensional embedding techniques. With a few exceptions, e.g., at magic numbers $N_\text{atoms}=59$, $65$, $76$ and $78$, we find that iron tends to form irregularly shaped NPs without a dominant surface character or characteristic atomic motif, and no reminiscence of crystalline features. We hypothesize that the observed disorder stems from an intricate balance and competition between the stable bulk motif formation, with bcc structure, and the stable surface motif formation, with fcc structure. We expect these results to improve our understanding of the fundamental properties and structure of low-dimensional forms of iron, and to facilitate future work in the field of iron-based catalysis.
\end{abstract}

%\keywords{}

\maketitle

\section{Introduction} % introduction
Iron \glspl{NP} are widely used for catalytic purposes, e.g., for the hydrogen evolution reaction (HER)~\cite{Cilpa-Karhu2019,Ahsan2020}, the oxygen reduction reaction (ORR)~\cite{Ahsan2020} or light olefin synthesis~\cite{TorresGalvis2012,Gu2020}.
To enable an economy not relying on crude oil for energy or base chemicals, the development of cost-effective and scalable catalysts for these reactions is crucial. Non-precious catalysts such as iron \glspl{NP} are of special interest because of their high reactivity and low cost compared to Pt- and Pd-based catalysts.

To study the atomic-level processes of these reactions in detail, while considering a wide variety of \glspl{NP} and active sites, an accurate but also computationally cheap model is necessary. However, among the existing models, \gls{DFT} is too expensive for such comprehensive studies and the different classical interatomic potentials lack in accuracy and general applicability.
Early on, many \gls{EAM} potentials were developed for general application to \gls{bcc} $\alpha$-Fe~\cite{Mendelev2003,FinnisSinclair,Dragoni2018EAM,Partay2018}. These were followed by a \gls{BOP}~\cite{Muller2007}, even able to describe magnetic interactions~\cite{Mrovec2011} and, to some extent, \gls{fcc} $\gamma$-Fe.
Later, potentials for special purposes were created, for instance to study radiation defects~\cite{Alexander2020} or the conditions in the Earth's core~\cite{Sun2022b}.
Recently, different flavors of \gls{ML} potentials~\cite{deringer_2019} have been trained for iron, including neural network potentials~\cite{Cian2021,Meng2021} and \glspl{GAP}~\cite{Dragoni2018GAP,Zhang2022}.

Some of the potentials mentioned above claimed general applicability, but are not truly general as they are not applicable to all crystal phases, surfaces, \glspl{NP} and disordered structures (including the liquid). Rather, they are typically designed to describe a wide range of properties of $\alpha$-Fe, with no guarantee of transferability outside of this range. In this work, we present a new \gls{GAP} \gls{MLP} trained on a much wider range of structures for true general applicability and transferability across a wide range of problems in atomistic modeling of iron.
The results are compared to the \gls{GAP} potential by Dragoni \etal~\cite{Dragoni2018GAP}, as the current state of the art, and the \gls{EAM} potential by Mendelev \etal~\cite{Mendelev2003}, as one of the most used classical alternatives with a lower computational cost.
These were chosen as representative of two different philosophies for deriving interatomic potentials: empirical potentials fitted to material properties and \glspl{MLP} trained on energies and forces computed from \gls{DFT}. While both types of potentials are ultimately designed to reproduce the material properties, which are in turn linked to the energies and forces of atomic configurations, the two approaches should be distinguished as they lead to different tradeoffs between accuracy and computational cost.

We showcase the ability of our \gls{GAP} to accurately describe the \gls{PES} of crystalline and nanostructured iron, including a reasonable description of different surfaces and phase transformations at extreme thermodynamic conditions. The highlight application of this paper is the search for stable iron \glspl{NP} of different sizes, a task for which our \gls{GAP} achieves accuracy remarkably close to that of \gls{DFT}. We hope that this work will speed up the discovery of efficient iron-based nanocatalysts.

\section{Database generation}
There are three critical steps in training a GAP: 1) training database generation, 2) selection of model architecture and hyperparameters, including the choice of atomic descriptors and data regularization, and 3) the computation of the fitting coefficients. These have been covered in detail in the literature, and we refer the reader to Refs.~\cite{bartok_2015,deringer_2021} for an in-depth discussion. Here, we will only give a brief account of the technical ingredients of our \gls{GAP} and focus mostly on accuracy benchmarks and applications.

Ensuring the accuracy and transferability of a \gls{GAP}, or any other \gls{MLP}, for that matter, relies on the availability of a database of atomic structures covering the relevant regions of configuration space. For a general-purpose \gls{MLP}, this means that comprehensive sampling of the \gls{PES} needs to be done. Our iron database contains dimers and trimers, crystalline structures (\gls{bcc}, \gls{fcc}, \gls{hcp}, \gls{sc} and diamond) over a wide range of cell parameters and with ``rattled'' atomic positions (i.e., atoms slightly displaced about their equilibrium positions), transitional structures between \gls{bcc}-\gls{fcc} and \gls{bcc}-\gls{hcp}, surface slabs cleaved from the relaxed bulk structures, \glspl{NP} and liquid configurations.
For each structure, the magnetic configuration with the lowest energy was chosen for inclusion in the database. In this way, our \gls{GAP} is fitted to the \gls{DFT} ground state with regard to the magnetic degrees of freedom, which are otherwise not explicitly taken into account in our \gls{PES} description.
Detailed types and numbers of structures in our training database are given in Table~S1 of the \gls{SM}~\cite{SM}.

The energy, forces and virials for the atomic structures in our training database were computed at the \gls{DFT} level of theory using VASP~\cite{Kresse1993,Kresse1996a,Kresse1996b}. We used the PBE functional~\cite{perdew_1996} with standard PAW pseudopotentials~\cite{bloechl_1994,Kresse1999} for Fe (with 8 valence electrons, $4s^2 3d^6$). The kinetic energy cutoff for plane waves was set to $400$~eV and the energy threshold for convergence was $10^{-7}$~eV. All the \gls{DFT} calculations were carried out with spin polarization, which can describe collinear magnetism. While non-collinear magnetic effects can in principle be described in VASP, the gain in accuracy in the context of \gls{MLP} simulation is only modest compared to the increased CPU cost and difficulty to systematically converge thousands of individual calculations in a high-throughput setting.

On this database, we trained our \gls{GAP} with $2$-body, $3$-body and many-body \gls{SOAP}~\cite{Bartok2013,Caro2019} atomic descriptors using a cutoff of $5$~\AA{}, $3$~\AA{} and $5$~\AA{}, respectively. A ``core'' potential, a tabulated pairwise interaction at very short interatomic distances, was added to model the strongly repulsive regime down to $0.1$~\AA{}.
The number of sparse configurations and the regularization parameter were both chosen per configuration type, and are listed in Table~S1 of the \gls{SM}~\cite{SM}. The training was carried out with the QUIP/GAP codes~\cite{csanyi_2007,ref_quip}. The full command passed to the \texttt{gap\_fit} binary is given in the \gls{SM}~\cite{SM}. We again refer the reader to the literature for further details on GAP training~\cite{bartok_2015,deringer_2021}.

\section{GAP validation}
In this section we validate our \gls{GAP} against a wide range of simulation problems and compare it to existing potentials. We first motivate the need and usefulness of a general-purpose \gls{MLP} for iron. We then benchmark the \gls{GAP} for the description of bulk iron, phase transitions, elastic properties and surface calculations.

\subsection{General-purpose vs bcc-specific iron potential}
While ferromagnetic \gls{bcc} is the ground-state structure of bulk iron at room temperature and pressure, iron transitions to other stable structures as the thermodynamic conditions change. In addition, surfaces cleaved from the bulk look different depending on the bulk crystal structure. Nanostructured iron, in particular \glspl{NP}, will not necessarily have a \gls{bcc} or, for that matter, ordered structure, even at room temperature and pressure. Finally, liquid iron is simply disordered and thus its structure differs significantly from \gls{bcc} or any other crystal structure. Interatomic potentials trained from \gls{bcc} data can be very useful to accurately describe the properties of $\alpha$-Fe, but their accuracy deteriorates rapidly as they extrapolate in regions of configuration space away from the training data. Here we show how our \gls{GAP} overcomes these issues and provides a consistent prediction of the \gls{PES} of iron for widely different problems, enabling an accurate description of \glspl{NP} of varying sizes. We will also show that, in the absence of an explicit inclusion of the magnetic degrees of freedom, this transferability is achieved at the cost of sacrificing accuracy in the description of some of the properties, e.g., of the surface energetics of the different crystal phases.

\begin{figure*}[t]
\begin{tabular}{c c}
RMSE on our training database (general) & RMSE on Dragoni \etal's database (bcc-specific) \\
\includegraphics[width=0.48\linewidth,keepaspectratio]{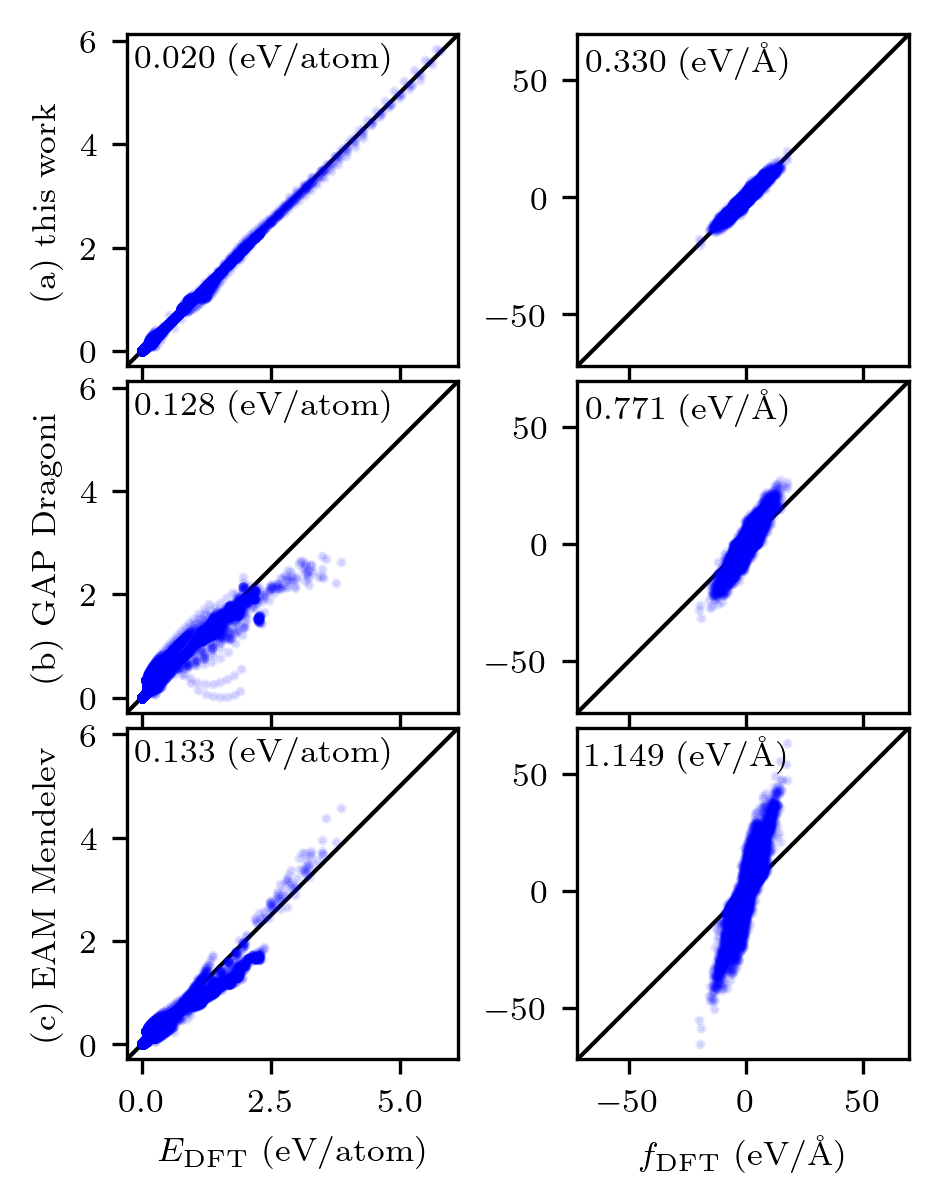} &
    \includegraphics[width=0.48\linewidth,keepaspectratio]{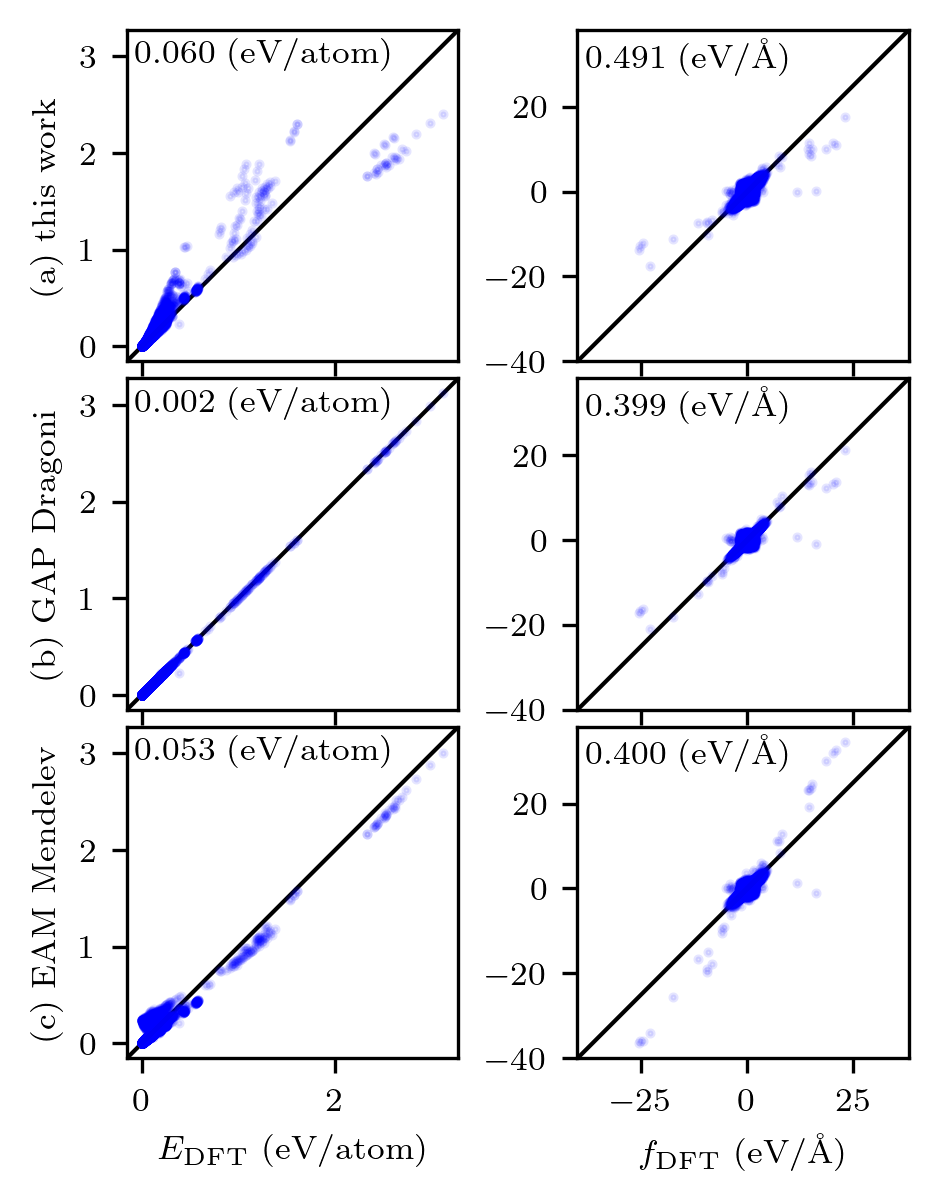}
\end{tabular}
    \caption{Energy and force \gls{RMSE} of (a) the GAP potential developed in this work, (b) the \gls{GAP} potential by Dragoni \etal~\cite{Dragoni2018GAP} and (c) the \gls{EAM} potential by Mendelev \etal~\cite{Mendelev2003} compared to \gls{DFT}. On the left are the \glspl{RMSE} computed on our training database and on the right are the \glspl{RMSE} on the training database of Dragoni \etal~\cite{Dragoni2018GAP}. The energy values have been referenced to the bulk energy of \gls{bcc} iron for each potential. Note that on the left, only the panel for our \gls{GAP} includes the diamond and \gls{sc} structures.}
    \label{fig:EnergyForce}
\end{figure*}

Figure~\ref{fig:EnergyForce} shows the energy and force errors of (a) the \gls{GAP} potential developed in this work, (b) the \gls{GAP} potential by Dragoni \etal~\cite{Dragoni2018GAP} and (c) the \gls{EAM} potential by Mendelev \etal~\cite{Mendelev2003} against the corresponding \gls{DFT} values in two \textit{training} databases: the training database of our \gls{GAP} on the left and that of the Dragoni \gls{GAP} on the right. The energy values have been referenced to the bulk energy of \gls{bcc} iron for each potential, to make the results comparable between the potentials. Each panel shows the \gls{RMSE} for the respective data. Importantly, \fig{fig:EnergyForce} is not intended as an accuracy test of our \gls{GAP}, for which reporting the training-set \gls{RMSE} is meaningless. We rather use it to showcase the difficulty encountered by bcc-specific iron potentials to reproduce the \gls{PES} of other structures, on the one hand, and to quantify the ability of the \gls{GAP} framework to learn the \gls{PES} within a significantly more comprehensive region of configuration space than that corresponding to \gls{bcc}, on the other. In this regard, our \gls{GAP} is able to learn our general-purpose training database to an accuracy of $20$~meV/atom which, while satisfactory for many purposes, is significantly higher than the $2$~meV/atom with which the Dragoni \gls{GAP} can learn its own \gls{bcc}-specific training database. At the same time, our \gls{GAP} only significantly deviates in the predictions of high-energy structures in the Dragoni database, with an overall \gls{RMSE} of $60$~meV/atom, mostly arising from outliers in the high-energy regions of the \gls{bcc} \gls{PES}, whereas the Dragoni \gls{GAP} struggles to capture the energetics of many low-energy structures in our database (as well as the high-energy ones), with an overall \gls{RMSE} of $128$~meV/atom. The \gls{EAM}'s performance is more predictable, with reasonably good \glspl{RMSE} for \gls{bcc} iron.

We note that the test in \fig{fig:EnergyForce} (left) was done only for physically meaningful structures. The diamond and simple cubic structures used in the training of our \gls{GAP} are not included in the plots for the reference potentials, as these structures are high in energy and not physically meaningful and would not make for a fair, nor instructive, comparison. They are included in the plot for our \gls{GAP}, though, increasing the \gls{RMSE} from $12$ to $20$~meV/atom there.
Dimer and trimer structures are excluded from the plots as well, as they reach very high energies and would obscure the more important data ranges. All other configurations used in the training are shown here, including the different bulk crystal structures, surfaces, melt, vacancies and \glspl{NP}.

Unsurprisingly, the energies in our training database are very well reproduced by the \gls{GAP} potential developed in this work. The GAP potential by Dragoni \etal~\cite{Dragoni2018GAP} reproduces the energies almost to the same \gls{RMSE} as the \gls{EAM} potential by Mendelev \etal~\cite{Mendelev2003}. For both potentials, nucleation clusters, \glspl{NP} and structures derived from \gls{hcp} pose the greatest problems, with \glspl{RMSE} of $312$, $171$ and $151$~meV/atom for the Dragoni \gls{GAP}, and $460$, $343$ and $119$~meV/atom for the Mendelev \gls{EAM}.
The energies of structures derived from \gls{bcc} on the other hand are predicted best, with \glspl{RMSE} of $8$ and $11$~meV/atom, respectively. 

Again, unsurprisingly, the forces in our training database are reproduced very well with the \gls{GAP} potential developed in this work. Both the \gls{GAP} potential by Dragoni \etal~\cite{Dragoni2018GAP} and the \gls{EAM} potential by Mendelev \etal~\cite{Mendelev2003} predict forces that are systematically too large by factors of approx. $1.1$ and $1.6$, respectively.

% overall energy / force errors vs. DFT
The energy and force errors shown in \fig{fig:EnergyForce} are significantly lower with our \gls{GAP} than with the reference potentials, which had to be expected as the data contains many types of structures that the reference potentials were never intended for.
It is still not a given that our \gls{GAP} would reproduce the energies and forces so well in the different regions of configuration space, as we have observed that the addition of training data in one region usually leads to a very slight degradation in the other regions, a phenomenon that we ascribe to the absence of explicit magnetic degrees of freedom in our \gls{GAP}.
Also note the slopes of the force error data for the two reference potentials, especially the Mendelev \gls{EAM}. This shows a typical behavior of empirical potentials to have too strong a driving force towards stable configurations, by design, in order to avoid unstable trajectories in \gls{MD} simulations.

\subsection{Description of bulk iron}
\begin{figure}[t]
    \includegraphics[width=\linewidth,keepaspectratio]{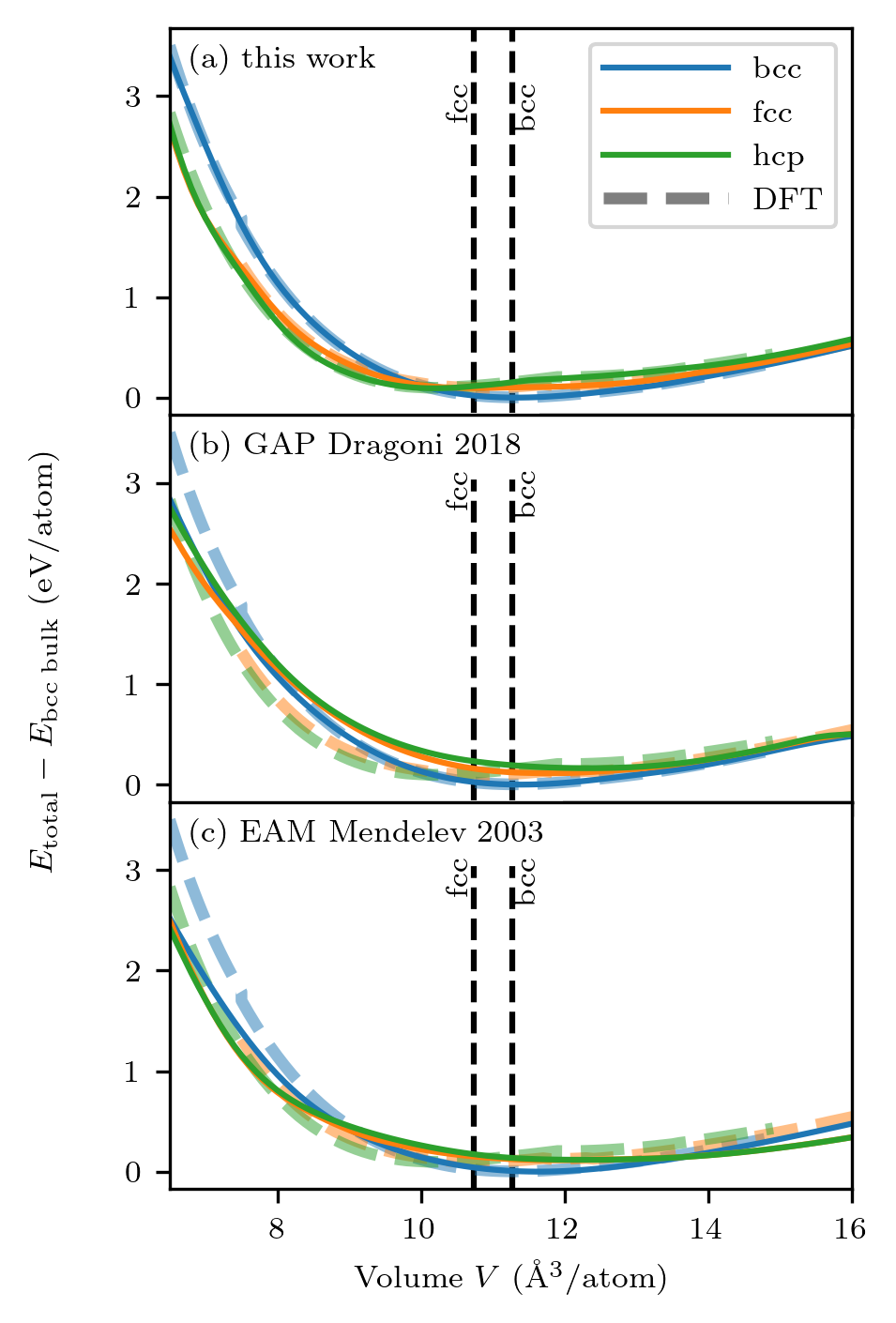} \\
    \caption{Energy of different Fe crystal phases over a wide range of atomic volumes: (a) the \gls{GAP}  of this work, (b) the \gls{GAP} potential by Dragoni \etal~\cite{Dragoni2018GAP}, (c) the \gls{EAM} potential by Mendelev \etal~\cite{Mendelev2003}. Vertical black lines mark the equilibrium atomic volumes of the bcc and fcc phases. DFT data for each crystal phase are underlayed as thick dashed lines in each panel.}
    \label{fig:Stability}
\end{figure}

Figure~\ref{fig:Stability} shows the relative stability of the bulk crystal phases vs. the atomic volume for all three potentials. Curves over a larger range of atomic volumes and additional diamond and \gls{sc} structures are shown in the \gls{SM}~\cite{SM}, Fig.~S1.
The shaded dashed curves represent \gls{DFT} reference values. The atomic volumes of the \gls{DFT} minimum structures for \gls{bcc} and \gls{fcc} are marked in each panel.

The \gls{GAP} developed in this work and the \gls{GAP} by Dragoni \etal{} capture the \gls{bcc} curve very well, including the minimum. The \gls{EAM} by Mendelev \etal{} has the minimum at a slightly too large volume and diverges from the \gls{DFT} reference data at lower and higher atomic volumes.
The \gls{fcc} energies are only reproduced well throughout the whole range considered by our \gls{GAP}. The \gls{GAP} by Dragoni \etal{} only gives the correct energies from approx. $11.5$ to $16$~\AA$^3$/atom. Towards lower atomic volumes the slope is much too steep, erroneously predicting \gls{fcc} iron to be less stable there than \gls{bcc} iron. The \gls{EAM} by Mendelev \etal{} does not capture \gls{fcc} well at all, with an exception around $12$~\AA$^3$/atom where it gives approximately the correct energies (but wrong trends).
However, there the \gls{hcp} energy is much too low, with both phases predicted as having similar energy. From $8$ to $10$~\AA$^3$/atom, where \gls{hcp} should be the stable crystal phase, either \gls{bcc} appears as more stable or all three crystal phases are practically identical in energy. With the Dragoni \gls{GAP}, \gls{hcp} is never the most stable structure in the atomic volume range shown here. At even larger volumes (see Fig.~S1 in the \gls{SM}~\cite{SM}), \gls{hcp} becomes more stable, but spuriously so, with a predicted energy almost as low as for the bulk \gls{bcc} minimum. The \gls{GAP} developed in this work predicts the correct energies also for the \gls{hcp} structure over the whole volume range studied.

Therefore, both the Dragoni \gls{GAP} and the Mendelev \gls{EAM} show strong deviations from the \gls{DFT} stabilities and predict the \gls{bcc} phase as the stable one over too wide a volume range, with the high-pressure \gls{hcp} phase missing. This can be easily attributed to the fact that the reference potentials were developed for the \gls{bcc} phase, neglecting the other crystal structures.
Of the problems at very high volumes (see Fig.~S1 of the \gls{SM}~\cite{SM}), only one seems important: the unphysical behavior of \gls{hcp} beyond 16~\AA{}$^3$/atom could lead to configurations blowing up during \gls{MD}. (This has been fixed in the fracture \gls{GAP}~\cite{Zhang2022}, based on the Dragoni \gls{GAP}, see Fig.~S2.)

\begin{figure}[t]
\centering
    \includegraphics[width=1\linewidth,keepaspectratio]{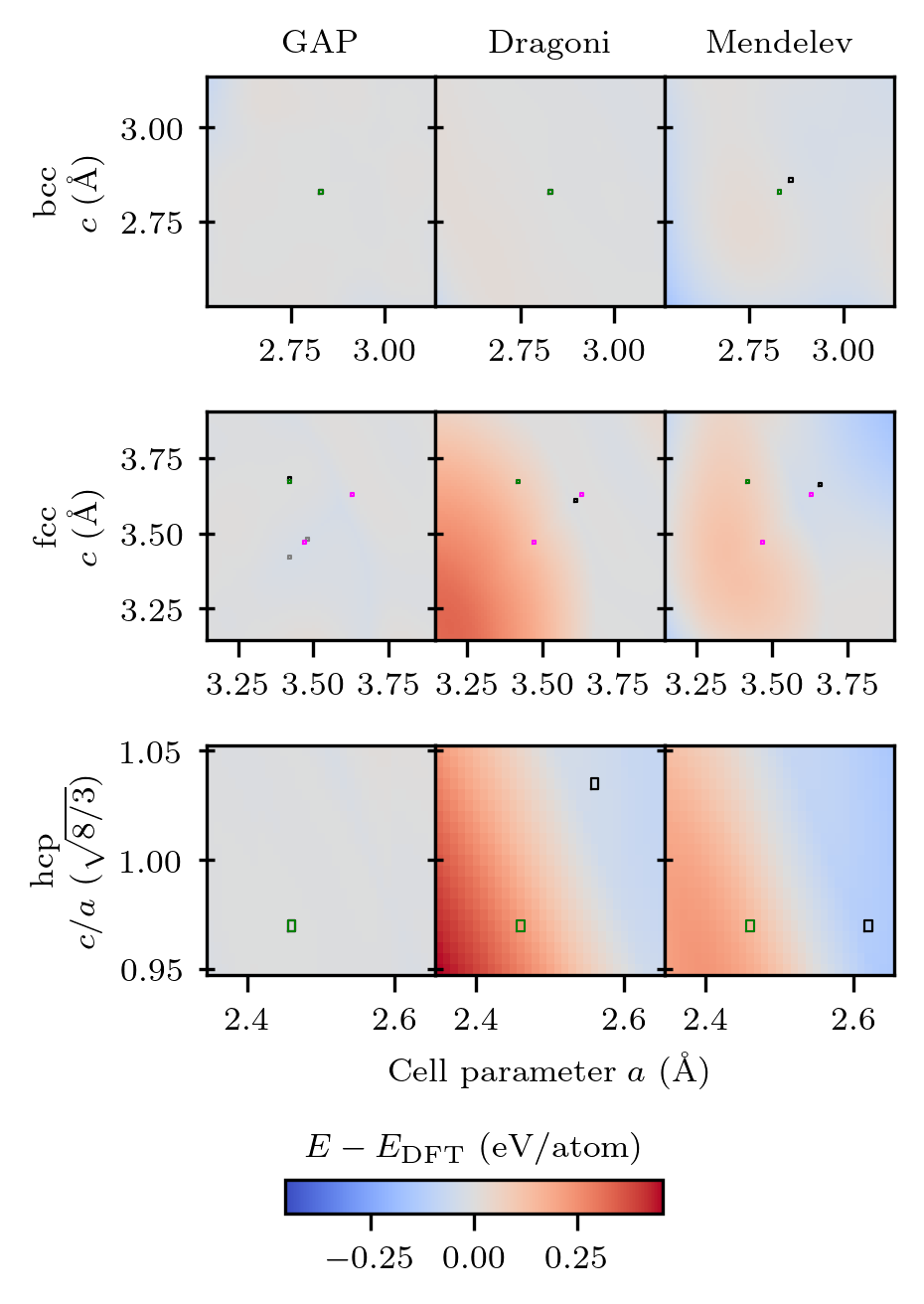}
    \caption{Energy error (difference between interatomic potential and DFT energies) over a space of structural parameters. Rows contain \gls{bcc}, \gls{fcc} and \gls{hcp} crystal structures and columns contain the three different potentials: the \gls{GAP} potential developed in this work, the \gls{GAP} potential by Dragoni \etal~\cite{Dragoni2018GAP} and the \gls{EAM} potential by Mendelev \etal~\cite{Mendelev2003}. The first structural parameter is always the cell parameter $a$ and the second is the cell parameter $c$ for \gls{bcc} and \gls{fcc}, but the aspect ratio $c/a$ for \gls{hcp}. Marked in black are the lowest-energy structures in each panel for the corresponding potential, marked in green the lowest \gls{DFT} energy structures. Interatomic potential and DFT data agree when green and black rectangles overlap on the graph. Additional local minima are marked in grey and purple for the interatomic potentials and \gls{DFT}, respectively.}
    \label{fig:CellParams}
\end{figure}

% crystal structures and cell parameters
Figure~\ref{fig:CellParams} shows the energies for strained bulk cells at different cell parameters for all three crystal structures and the three different potentials, each as the difference to the corresponding \gls{DFT} value.
For \gls{bcc}, the energy landscape is reproduced equally well by the \gls{GAP} potential by Dragoni \etal~\cite{Dragoni2018GAP} and the \gls{GAP} developed in this work. The \gls{EAM} potential by Mendelev \etal~\cite{Mendelev2003} shows larger disagreement with the \gls{DFT} reference, especially at low atomic volumes.
Both for \gls{fcc} and \gls{hcp}, the energy error is significantly lower for the \gls{GAP} developed in this work compared to the other two potentials. Noticeably, the Mendelev \gls{EAM} performs slightly better than the Dragoni \gls{GAP}. Both potentials overestimate the energy at low and underestimate it at high atomic volumes.

The lowest-energy structure is marked in black in each panel of \fig{fig:CellParams} for the respective potential (\gls{FM} \gls{bcc}, \gls{AFM} \gls{fcc}, \gls{NM} \gls{hcp}). The lowest DFT energy structure is marked in green in all panels for reference. Local minima with higher energy are marked in grey and purple for the interatomic potentials and \gls{DFT}, respectively.
The minima for the \gls{GAP} developed in this work coincide with the \gls{DFT} minima for \gls{bcc} and \gls{hcp}.
For the Dragoni \gls{GAP} the \gls{bcc} minimum differs only marginally from the \gls{DFT} reference, for the Mendelev \gls{EAM} slightly more.
The \gls{hcp} cell parameters predicted by the Dragoni \gls{GAP} and the Mendelev \gls{EAM} deviate substantially from the \gls{DFT} reference.

For the \gls{fcc} cell, multiple minima exist for \gls{DFT}~\cite{Muller2007,Herper1999}. The lowest in energy is the \gls{AFM} magnetic configuration with a tetragonal cell ($c$ longer than $a$), but two \gls{FM} minima with cubic cells exist as well, usually called \gls{FMLS} and \gls{FMHS}.
Of the three potentials, only the \gls{GAP} developed in this work reproduces more than one minimum structure correctly: the \gls{AFM} and the \gls{FMLS}.
The Dragoni \gls{GAP} does have a cubic and a tetragonal minimum structure  as well, but with $a=3.416$~\AA{} and $c=4.042$~\AA{}, the tetragonal minimum is outside of the plotting range of \fig{fig:CellParams}.
The Mendelev \gls{EAM} does not have a tetragonal cell minimum at all, but just a cubic cell minimum corresponding to the \gls{FMHS} at approx. the correct cell parameter, with cell parameter slightly too large at $a=3.658$~\AA{}, compared $a=3.634$~\AA{} from \gls{DFT}.
All cell parameters are given in Table~S2 in the \gls{SM}~\cite{SM}.

% cell parameters
The reproduction of the cell parameters of the crystalline phases shown in \fig{fig:CellParams} works well for the \gls{bcc} phase with all three potentials. Still, for the Mendelev \gls{EAM} the deviation from the \gls{DFT} cell parameters and the errors in the surrounding energy landscape are larger than with the Dragoni \gls{GAP} and the \gls{GAP} developed in this work.
However, for the \gls{fcc} and \gls{hcp} phases, both reference potentials yield large errors in the energy landscape and cell parameters far from the \gls{DFT} ones, while our \gls{GAP} gives very low errors and the correct cell parameters.
While multiple local minima exist for the cell parameters of the \gls{fcc} cell, one for the tetragonal \gls{AFM} configuration and two for the cubic \gls{FMLS} and \gls{FMHS} states, no potential has a minimum for all three of these.
Thus, for our \gls{GAP}, the only shortcoming here is that it does not have a minimum for the \gls{FMLS} structure. The reason for this is the way that \gls{GAP} fits the underlying data smoothly, potentially removing shallow minima in some instances. Note that, even if the \gls{FMHS} minimum is missing, the error is still low.

We remark here that the inability of the Dragoni \gls{GAP} and Mendelev EAM potentials to accurately describe crystal phases other than \gls{bcc} is to be expected, and not an artifact, since they were designed to correctly describe \gls{bcc} iron only. An accurate description of the low-pressure \gls{bcc} structures can still be obtained with these potentials, especially with the Dragoni \gls{GAP}, which should be able to indeed outperform our GAP for simulation of single-phase \gls{bcc} iron.

\subsection{Phase transitions}
\begin{figure}[t]
\centering
    \includegraphics[width=\linewidth,keepaspectratio]{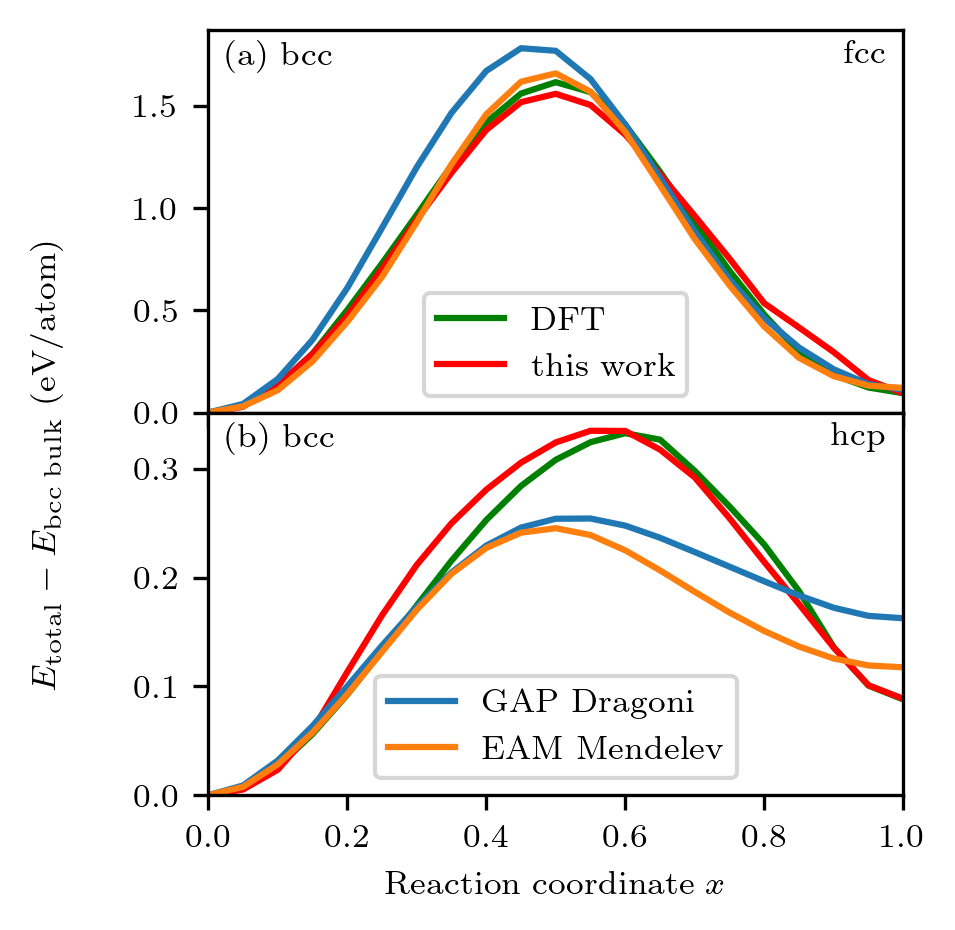}
    \caption{Transition between (a) bcc and fcc and (b) bcc and hcp crystal structures calculated with \gls{DFT}, the \gls{GAP} potential developed in this work, the \gls{GAP} potential by Dragoni \etal~\cite{Dragoni2018GAP} and the \gls{EAM} potential by Mendelev \etal~\cite{Mendelev2003}. Energies are referenced to the bcc bulk energy for each potential.}
    \label{fig:Transitions}
\end{figure}

So far we have discussed the accuracy of our \gls{GAP} to describe (meta)stable structures. However, a general-purpose potential to be used in dynamic structure generation, e.g., involving \gls{MD} simulation, also needs to accurately describe the \gls{PES} along important transition paths between crystal phases.
Since the initial and final states, as well as the minimum energy path for a transition, depend on the specific force field used, we choose the following approach to be able to compare the potentials to DFT and among them.
For the transformations from \gls{bcc} to \gls{fcc} and from \gls{bcc} to \gls{hcp}, transition structures were created by \textit{linear interpolation} between the cell parameters and atomic positions of the endpoint structures at $19$ points along the path, as shown in \fig{fig:Transitions}. The minimum \gls{DFT} energy structures were used for all potentials, i.e., to the two reference potentials these endpoints are not the minimum energy structures, but for our \gls{GAP} they are, as there the cell parameters are identical to the \gls{DFT} ones. Thus, at the \gls{fcc} and \gls{hcp} endpoints, the Dragoni and Mendelev energies differ from the \gls{DFT} reference values.
All curves are referenced to the \gls{bcc} bulk energy.

Along the \gls{bcc} to \gls{fcc} path shown here, our \gls{GAP} and the Mendelev \gls{EAM} reproduce the energy barrier reasonably well, our \gls{GAP} a little too low and the Mendelev \gls{EAM} a little too high. The Dragoni \gls{GAP} energies along the path are considerably higher than with the other potentials.
Along the \gls{bcc} to \gls{hcp} path, the Dragoni \gls{GAP} and the Mendelev \gls{EAM} trace the \gls{DFT} curve up to $x \sim 0.3$ (where $x$ is the reaction coordinate), but then quickly diverge and reach an endpoint far from the \gls{DFT} one (more so for the Dragoni \gls{GAP} than the Mendelev \gls{EAM}).
Our \gls{GAP}'s curve has a slightly different shape, with a steeper incline at low $x$ and a higher maximum, but still fits the \gls{DFT} curve much better than the other two potentials.

We note again that these curves were not obtained through the minimum energy path, e.g., by performing a \gls{NEB} calculation~\cite{jonsson_1998}, but rather by linear interpolation between the endpoint structures. Hence, the maxima in the transition curves cannot be interpreted as ``barriers'' (i.e., the energy calculated at the saddle point along the minimum energy path), as lower-barrier paths might exist.

\subsection{Elastic properties}
\begin{figure}[t]
\centering
    \includegraphics[width=1\linewidth,keepaspectratio]{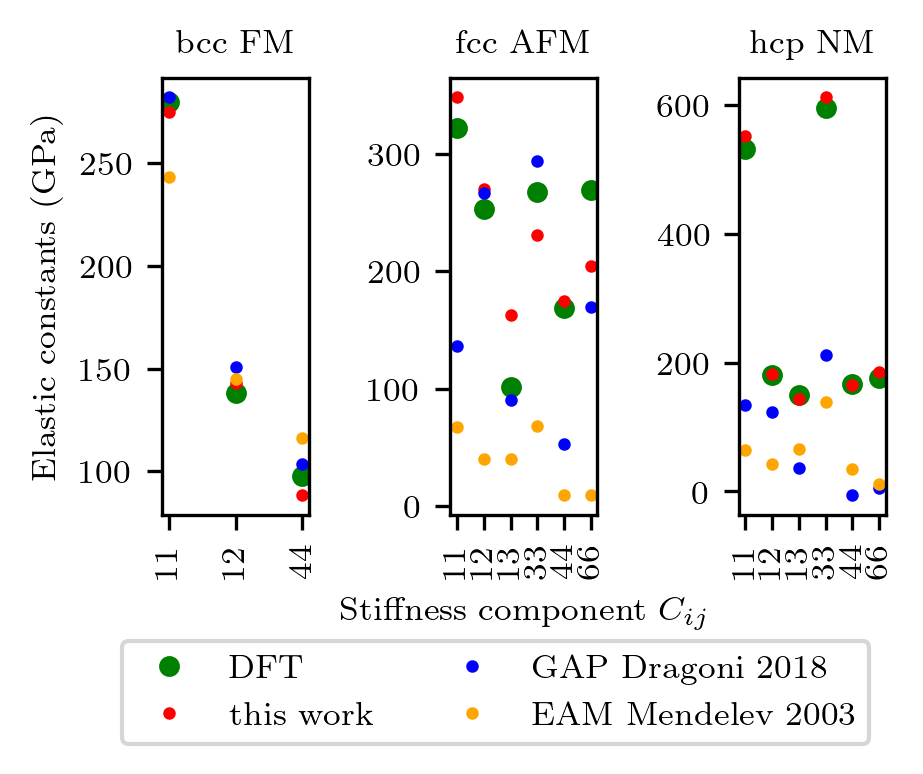}
    \caption{Elastic constants for all crystal structures and potentials, compared with \gls{DFT} results. Note that the Mendelev \gls{EAM} predicts a cubic symmetry for the \gls{fcc} structure, that is predicted tetragonal by \gls{DFT}.}
    \label{fig:Elasticity}
\end{figure}

The elastic constants of all three \gls{DFT} minimum-energy structures have been computed with all potentials by straining the structures in all relevant directions (depending on the symmetry) in $5$ steps with strain increments in the range of $10^{-5}$ to $10^{-2}$ to check for consistency, as implemented in \gls{ASE}~\cite{Xiao2017}. The results are shown in \fig{fig:Elasticity}.

The elastic constants of the \gls{bcc} structure are generally reproduced well by both the \gls{GAP} by Dragoni \etal{} and our \gls{GAP}. With the Mendelev \gls{EAM}, $C_{11}$ is noticeably too low by approx. $10\%$ and for $C_{44}$ the difference to the \gls{DFT} reference value is about twice as large as for the two \gls{GAP} potentials. The error for $C_{12}$ is only half as large as with the Dragoni \gls{GAP} potential, though.
For the \gls{hcp} elastic constants, our \gls{GAP} also yields low errors with respect to the \gls{DFT} data. Both the reference potentials significantly underestimate the elastic constants. For the \gls{fcc} elastic constants, the results are much more mixed: while most of the Mendelev \gls{EAM} values are significantly too low, our \gls{GAP} and the Dragoni \gls{GAP} give values that are a mix of too low or high ones with some that are spot on.

For the elastic constants shown in \fig{fig:Elasticity}, again only the \gls{bcc} phase is represented well by all three potentials, with the Mendelev \gls{EAM} showing the largest errors compared to the \gls{DFT} reference values.
However, the elastic constants of the \gls{hcp} phase are only reproduced well by our \gls{GAP} and the \gls{fcc} phase is not reproduced well by any of the potentials. While the reference potentials were not developed for these crystal structures, our \gls{GAP} does have the necessary structures in the training database and still fails in the prediction for the \gls{fcc} phase.
Although we tried extensively to train a GAP that could correctly reproduce the elastic constants of all three crystal phases by fine tuning the regularization parameters and sparse set configurations of the strained+rattled structures, we did not manage to obtain a fit that predicted all of them accurately at the same time. We attribute this to the fact that each crystal structure belongs to a different magnetic branch with possibly significantly different energetics, including the energy derivatives (i.e., forces and the stress tensor, used to compute the elastic constants). 
Different magnetic configurations exist for the \gls{fcc} phase, depending on the stained state. Without the explicit treatment of the magnetic moments, the underlying energy landscape has discontinuities where the lowest energy magnetic state changes. Our GAP can resolve these branches, but only implicitly, whenever the structures are sufficiently different in terms of atomic arrangements.
While it should be possible to train a dedicated potential to reproduce the elastic constants of any one crystal phase and magnetic configuration, predicting all of them accurately with a general potential seems impossible within our current methodological framework, especially when many other types of configurations are also considered. We speculate that only an iron \gls{MLP} which explicitly accounts for the magnetic structure of iron will be able to accurately capture all of these features simultaneously. Augmenting the \gls{GAP} framework to incorporate magnetism is far beyond the scope of this work, but we expect advances in this area within the next few years.

\subsection{Surfaces}
Another stringent test for an interatomic potential is the prediction of surface energies cleaved and reconstructed along various crystallographic planes, since these structures look significantly different from the bulk.
To calculate the energies of surfaces with various Miller indices cleaved from the \gls{bcc}, \gls{fcc} and \gls{hcp} bulk structures, slabs with a number of layers between $4$ and $16$ were set up using \gls{ASE}~\cite{Xiao2017}. These were then relaxed using \gls{DFT} calculations with a fixed box size. We ensured that the amount of added vacuum perpendicular to the slabs was sufficient to allow for relaxation in this direction and rule out any interaction between the periodic copies of the slabs.
We observed these slabs to have nontrivial magnetic structure, e.g., showing a strong dependence on the number of atomic monolayers. Thus, converging these \gls{DFT} relaxations was not possible in every case and, even when the calculations converged, local energy minima were found that were not necessarily also the global minima.

The result of the \gls{DFT} relaxations is primarily dependent on the setup of the initial magnetic moments and the final magnetic configuration resulting from those. Large differences up to $\sim 100$~meV/atom for the same slab with different magnetic configurations were found. The \gls{fcc} surface slabs proved more problematic than those of the other crystal structures in this regard.
Typically, the \gls{fcc} magnetic configurations consisted of layers with opposite local magnetic moments (not necessarily all with the same magnitude), one or more atomic layers thick.

To extract meaningful surface energies from the energies of the slabs, convergence of the surface energy with respect to the slab thickness would be expected. This was found for many surface indices, but in some cases no convergence could be observed, indicating that only local minima were found for at least some of the slabs.
To improve the performance of our \gls{GAP} for surfaces, surface slabs relaxed with \gls{DFT} were further relaxed with earlier versions of our \gls{GAP} and single-point \gls{DFT} calculations of the resulting structures fed back into the training set in an iterative manner (known as ``iterative training''~\cite{deringer_2017}).

The surface structures in our database were split into two categories: thin slabs that are too thin to have a bulk-like region in the center and thicker slabs that do contain such a bulk-like center. For the thin slabs, the regularization parameter during training was chosen $25$ times higher than the default (the higher the regularization parameter the less stringently the GAP is required to follow the data). This was done to keep the structures in the database, but focus on the more realistic surfaces from the thicker slabs.

\begin{figure}[t]
\centering
    \includegraphics[width=1\linewidth,keepaspectratio]{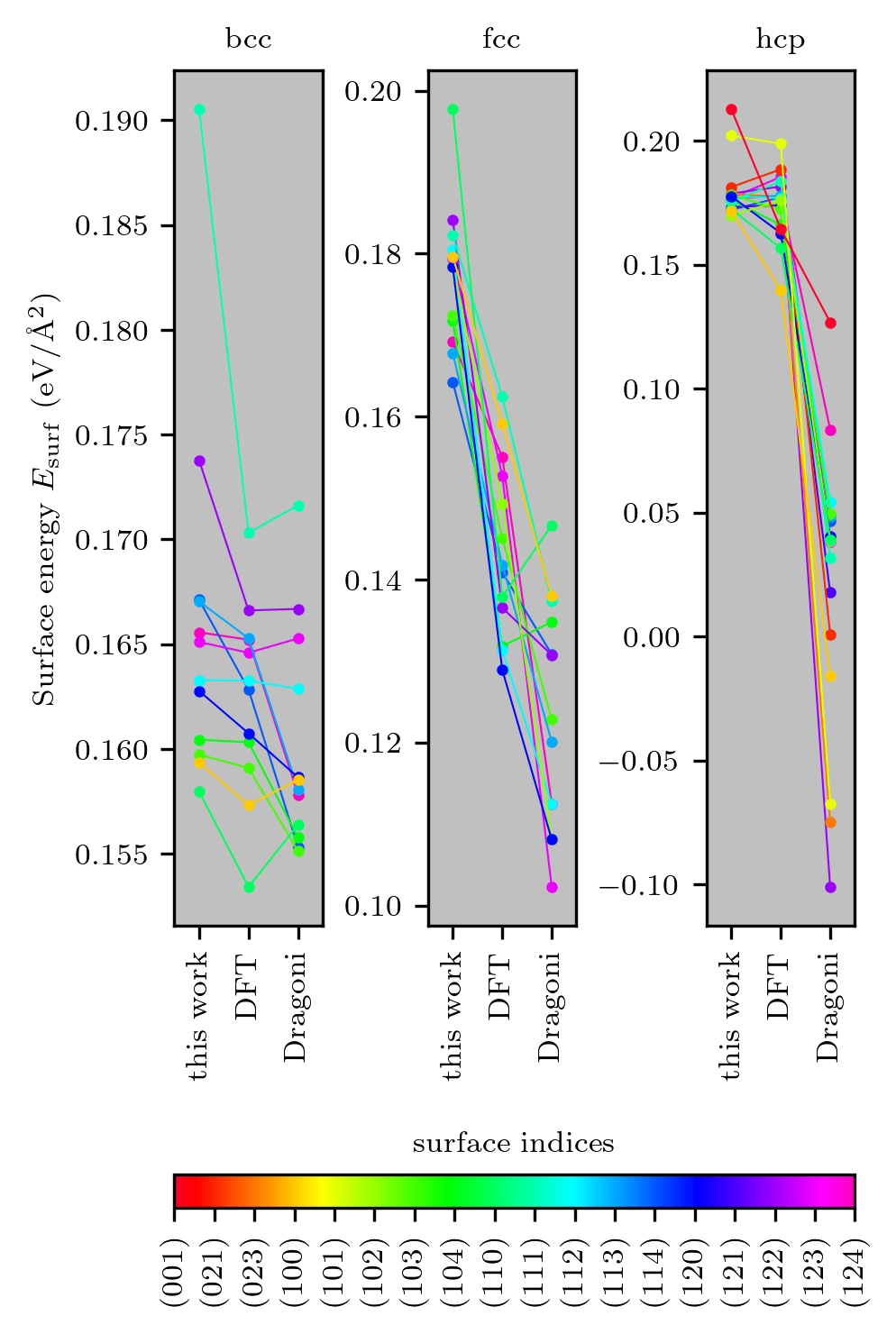}
    \caption{Surface energies of a wide range of surface indices for the \gls{GAP} by Dragoni \etal{} and our \gls{GAP}, compared with the \gls{DFT} values.}
    \label{fig:SurfaceEnergies}
\end{figure}

The surface energies for all three crystal structures and various surface indices are shown in \fig{fig:SurfaceEnergies}, for our \gls{GAP} and the \gls{GAP} by Dragoni \etal, compared against the \gls{DFT} values. Data points for a given surface index are connected by lines between the different potentials and the indices are encoded in color for easier tracking. An equivalent plot comparing our \gls{GAP} to the \gls{EAM} by Mendelev \etal{} is shown in Fig.~S3 of the \gls{SM}~\cite{SM}.

For the \gls{bcc} surface energies, both Dragoni's and our \gls{GAP} give results that are comparable and match the \gls{DFT} references reasonably well. While the overall value range for the Dragoni GAP better matches the one found using \gls{DFT} (especially for high-energy surfaces, which our GAP overestimates), our \gls{GAP} performs better in terms of reproducing the correct order of the low-energy surface indices, i.e., there are fewer crossings of the connecting lines.
For the \gls{fcc} surfaces, both \gls{GAP}s predict the \gls{DFT} reference values poorly. While the energies come out too high with our \gls{GAP}, the one by Dragoni shows energies that are too low. Recall the issues already mentioned with convergence of \gls{fcc} slab calculations with \gls{DFT} as a function of the number of atomic layers; this is likely a problem that necessitates the explicit inclusion of magnetic structure and cannot be fully solved within the standard \gls{GAP} framework.

The \gls{hcp} surface energies are predicted much too low with the Dragoni \gls{GAP}, for some indices even negative, a clear sign of extrapolation outside of the training set. Our \gls{GAP}, on the other hand, predicts all surface energies in a range similar to the one predicted by \gls{DFT}, although with sizable errors.
The \gls{EAM} by Mendelev \etal{} gives bad predictions for the surface energies of all three crystal structures (Fig.~S3 of the \gls{SM}~\cite{SM}).

The calculation of the surface energies shown in \fig{fig:SurfaceEnergies} was challenging for some of the surface indices, as finding the magnetic ground state for these systems is much harder than for small bulk unit cells. In a few instances, it was not even possible to converge the \gls{DFT} calculation at all. In other cases the magnetic layering changed erratically between slabs with a different number of layers, so that no convergence of the surface energy with increased slab thickness could be found. There, even trying to use the magnetic configuration of one slab to inform the initial configuration for a similar slab failed. This and the interpolation between different magnetic states in the fitting of our \gls{GAP} led to mixed results. Consequently, the surface energies are more accurate for \gls{bcc} where only the \gls{FM} state exists and fewer convergence problems occurred in the creation of the training database.
Still, our \gls{GAP} performs noticeably better than the other two potentials, giving a reasonable range of values for all surface energies and the correct order of the lowest-energy surfaces for \gls{bcc}.
The Dragoni \gls{GAP} performs only slightly worse than our \gls{GAP} for \gls{bcc} (where it was trained), but gives even negative surface energies for \gls{hcp}, which could lead to instabilities in dynamics simulations.
Lastly, the Mendelev \gls{EAM} yields high errors for the surface energies of all crystal structures, as it was fitted only to properties of the crystalline bulk and liquid.

\section{Application to selected problems}
In this section we benchmark our Fe \gls{GAP} with representative use cases. We go beyond simple numerical scores, like \gls{RMSE} and \gls{MAE}, and focus on how the potential performs when trying to reproduce experimental trends in 1) thermal expansion, 2) the solid-liquid phase transition and 3) the temperature-pressure phase diagram.

\subsection{Thermal expansion}
The thermal expansion was studied using \gls{MD} calculations with \gls{ASE}~\cite{Xiao2017,Kermode2020-wu}. Systems with $1024$ atoms were set up in the minimum \gls{DFT} \gls{bcc} structure and kept initially at 200~K (the lowest temperature in the series) for $1$~ps to equilibrate and for another $2$~ps for averaging with a time step of $1$~fs. This procedure was repeated, step by step, increasing the temperature in intervals of $200$~K  up to $1600$~K.
Detailed simulation parameters are specified in the SM~\cite{SM} (see also references \cite{Melchionna1993,Melchionna2000} therein).

The data for the Dragoni \gls{GAP} lines up almost perfectly with their \gls{DFT} data~\cite{Dragoni2015}, which is itself at lower volume than the experimental data by, e.g., Basinski \etal~\cite{Basinski1955} and Ridley \etal~\cite{Ridley1968}.
The Mendelev \gls{EAM} comes closest to the experimental data, but only really agrees around $200$~K and between $1190$ and $1660$~K (where Basinski \etal{} found \gls{fcc} Fe, as opposed to the \gls{bcc} Fe predicted by \gls{EAM}).
Our \gls{GAP} predicts the lowest atomic volumes at all temperatures, lower than the \gls{DFT} data reported by Dragoni \etal, consistent with our \gls{DFT} data. We note the two discontinuities in experimental data by Basinski \etal{} that take place first at the $\alpha$ to $\gamma$ and then at the $\gamma$ to $\delta$ phase boundaries. The only one of these happening where simulation data is available ($\alpha$ to $\gamma$) is not captured by any of the potentials.

\begin{figure}[t]
\centering
    \includegraphics[width=\linewidth,keepaspectratio]{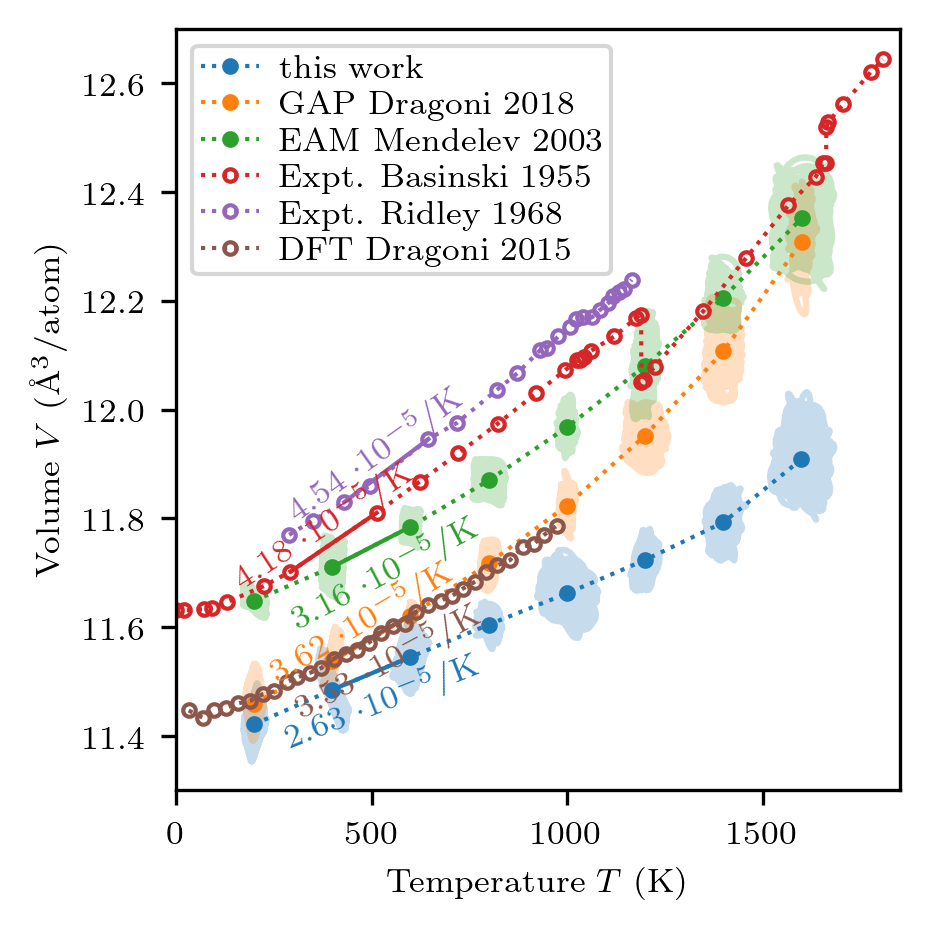}
    \caption{Atomic volume over a wide range of temperatures to show the thermal expansion. Experimental results~\cite{Basinski1955,Ridley1968} and \gls{DFT} calculations~\cite{Dragoni2015} shown as empty circles. For the calculations done in the scope of this work, volume and temperature along the trajectories are shown below the average values (full circles). For all curves, the coefficients of thermal expansion were fitted over the temperature ranges marked with solid lines. The coefficients are noted next to the curves.}
    \label{fig:thermal_exp}
\end{figure}

The coefficients of thermal expansion were fitted around $400$ to $600$~K, where data points were available. The fitting ranges are shown in \fig{fig:thermal_exp} as solid lines and the coefficients are noted next to the curves.
From the three potentials studied, the Dragoni \gls{GAP} comes closest to the experimental values and our \gls{GAP} differs the most. All three potentials underestimate the experimental values.

The thermal expansion curve for our \gls{GAP} starts at lower atomic volume than the reference potentials, in accordance with the cell parameters shown in \fig{fig:CellParams}. 
Also note the much larger sampled range of temperatures and volumes around the $1600$~K data point for our \gls{GAP}. We attribute this to the fact that the potential is exploring the energy landscape above the melting point (which is underestimated by our GAP, compared to experiment), but is missing a nucleation center for the liquid phase. We deal with the solid-liquid phase transition in Sec.~\ref{sec:solid-liquid}.

\subsection{Solid-liquid phase transition}\label{sec:solid-liquid}
The melting temperature of bcc iron was calculated using the two-phase method~\cite{Morris1994,Morris2002} with all three potentials. Systems with $\approx 11000$~atoms were set up as \gls{bcc} crystals and heated to $4000$~K for $15$~ps to melt half of the system, while the positions of the other half were kept fixed.
These half crystalline and half molten systems were then run at different target temperatures for $100$~ps of \gls{MD} with a time step of $1$~fs to find the temperature where both phases coexisted.
\gls{MD} simulations were carried out using LAMMPS~\cite{Plimpton1995} for the Mendelev \gls{EAM}~\cite{Mendelev2003} and Dragoni \gls{GAP}~\cite{Dragoni2018GAP} reference potentials and TurboGAP~\cite{Caro2019} for the \gls{GAP} developed in this work.
In the LAMMPS calculations, temperature and pressure were controlled with the Nos\'e-Hoover~\cite{Nose1998,Hoover1985} thermostat and barostat with damping constants of $1$~ps and $2$~ps, respectively. In TurboGAP, a Berendsen~\cite{Berendsen1998} thermostat and barostat with the same damping constants and a \texttt{gamma\_p}~\footnote{In TurboGAP, the bulk modulus for the barostat is expressed in units of the inverse compressibility of liquid water. E.g., \texttt{gamma\_p}~$= 55$ means that the material is assumed to be 55 times as incompressible as liquid water for the purpose of barostating. This allows the user to provide an intuitive value for this parameter whenever the compressibility factor of the system is not know \textit{a priori} (as is usually the case).} of $55$ were used.

\begin{figure}[t]
\centering
    \includegraphics[width=1\linewidth,keepaspectratio]{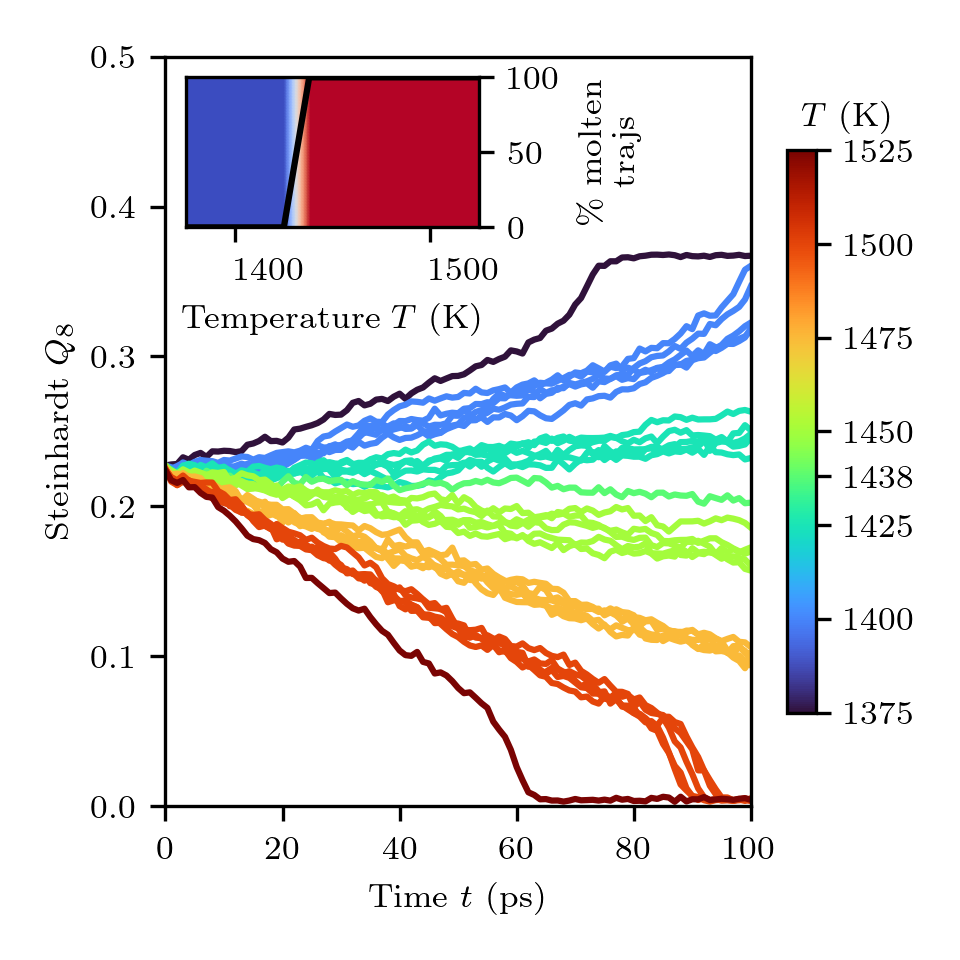}
    \caption{Steinhardt parameter $Q_8$~\cite{SteinhardtPaulJNelson1983} over time for \gls{MD} simulations at various temperatures in the range of $1400-1550$~K, around the melting temperature $T_m$. The inset shows the transition from the crystallizing to the melting state as a function of temperature.}
    \label{fig:Tmelt}
\end{figure}

Crystalline and molten states in the systems were identified with the Steinhardt parameter $Q_8$~\cite{SteinhardtPaulJNelson1983}. Figure~\ref{fig:Tmelt} shows the results for our \gls{GAP}, each line indicating a separate \gls{MD} run. (For the results using the Mendelev \gls{EAM} potential, see Fig.~S5 in the \gls{SM}~\cite{SM}.) Decreasing values of $Q_8$ indicate a melting system, while increasing values indicate a crystallizing one.
This gives approximate melting temperatures of $1760$~K and $1438$~K for the Mendelev \gls{EAM} and our \gls{GAP}, respectively.

For our \gls{GAP}, this is well below the experimental value of $1811$~K. (The experimental value is for the $\delta$ phase, which is also \gls{bcc} as in our simulations.) We attribute this deviation to the properties of the \gls{DFT} functional used for the training, which has in previous works been shown to predict too low melting points.
Part\'ay computed the phase diagram for the Mendelev \gls{EAM} using the nested sampling method~\cite{Partay2018}, believed to be the most comprehensive and accurate method for this purpose. She found a melting temperature higher than the experimental value, at approx. $1810-1940$~K.
The value of approx. $1760$~K we find for the Mendelev \gls{EAM} is right in the range of $1750-1775$~K they give in their original paper, somewhat in disagreement with the nested sampling result.
This disagreement can be attributed to a finite-size effect in the nested sampling calculations, overestimating the temperature of the melting transition compared to coexistence simulations.

For the Dragoni \gls{GAP} we found that the trajectories expanded to very high atomic volumes (about three times the volumes found with the other potentials for two-phase systems at the same temperature) upon releasing the crystalline atoms, immediately melting the crystalline half of the box. Thus, we were not able to stabilize the two-phase state at any temperature.
We attribute this to a spurious local minimum in the \gls{PES} for a low-pressure melt, lower than the pressurized crystal.

We note that our \gls{GAP} was not trained for thermal properties specifically, while the Dragoni \gls{GAP} was trained on data for thermomechanical properties and the Mendelev \gls{EAM} was fitted to the pair correlation function at $1820$~K. It is therefore not surprising that the Mendelev \gls{EAM} predicts the melting temperature better than the other two potentials.

\subsection{Phase diagram}
To compute the phase diagram we carried out free-energy calculations within the 2PT framework~\cite{lin_2003} as implemented in the DoSPT code~\cite{caro_2016,caro_2017b,ref_dospt}. 2PT computes the free energy of an ensemble of atoms from the integral of the density of states (calculated from \gls{MD}), which is partitioned between solid-like and gas-like degrees of freedom. This method is particularly suited to estimate the thermodynamic properties of liquids. In this work, we use it both for the liquid and the solid to be able to directly compare the free energy of the two and draw the melting curve: at any given set of thermodynamic conditions, the phase with the lowest free energy is the stable phase.

\begin{figure}[t]
\centering
    \includegraphics[width=1\linewidth,keepaspectratio]{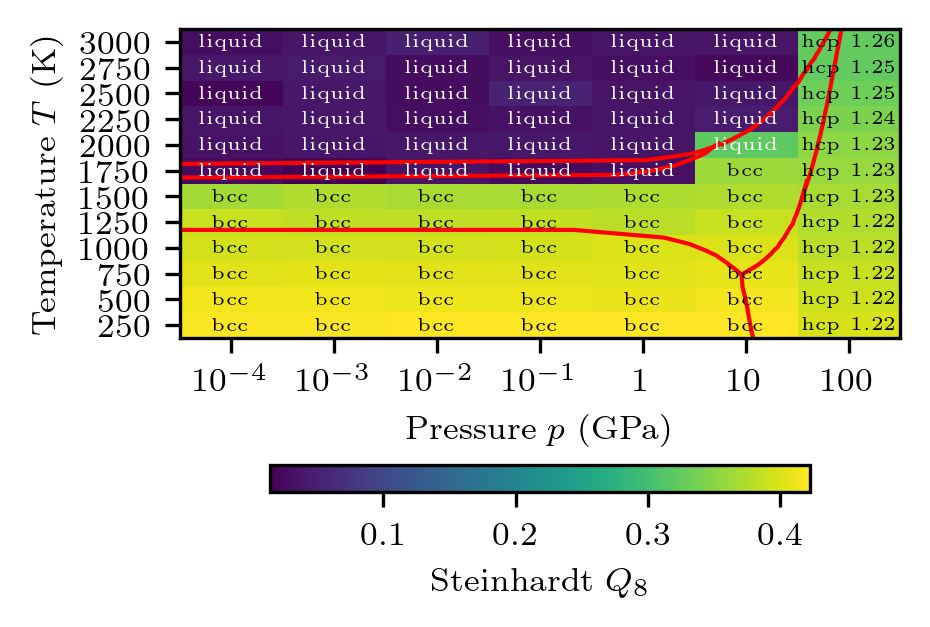}
    \caption{Phase diagram for our \gls{GAP} up to high temperatures and pressures. Crystalline structures were identified using \gls{XRD} spectra and by comparing \gls{SOAP} descriptors. For the \gls{hcp} phase, the $c/a$-ratio is given. Encoded in the color is the Steinhardt parameter $Q_8$~\cite{SteinhardtPaulJNelson1983}, to further highlight the solid-liquid transition. Red lines indicate the phase boundaries in the experimental phase diagram~\cite{Partay2018,Morard2018}: $\alpha$, $\gamma$, $\delta$ (in increasing $T$, at low $p$) and $\epsilon$ (at high $p$).}
    \label{fig:phase_diagram}
\end{figure}

We calculated the iron phase diagram with our \gls{GAP} up to high pressures of $100$~GPa ($10^6$~bar) and temperatures of $3000$~K, shown in \fig{fig:phase_diagram}. At each pressure, three \gls{MD} trajectories were initialized from $250$~K as \gls{bcc}, \gls{fcc} and \gls{hcp} and one from $3000$~K as liquid. The temperature was then increased/decreased in steps of $250$~K using the Bussi thermostat~\cite{Bussi2007} while controlling the pressure with the Berendsen barostat~\cite{Berendsen1998}. The calculations were done in TurboGAP~\cite{Caro2019} with equilibration constants of $100$~fs and $1000$~fs, respectively, and a \texttt{gamma\_p} of $100$. At each point, the trajectories were equilibrated for $80$~ps and subsequently sampled for $80$~ps.
Liquid structures were detected using the Steinhardt $Q_8$ parameter~\cite{SteinhardtPaulJNelson1983}, with low values indicating the liquid. We detected the crystalline structures by both comparing \gls{SOAP} descriptors and \gls{XRD} spectra calculated using the Debyer software package~\cite{ref_debye}. Example spectra at $p = 10^{-4}$~GPa ($1$~bar) and for the reference structures are shown in Fig.~S6 of the \gls{SM}~\cite{SM}.
For the \gls{hcp} structures at $100$~GPa the $c/a$-ratio is given in \fig{fig:phase_diagram}, determined by comparing to the \gls{SOAP} descriptors of \gls{hcp} with $c/a$-ratios in the range of $1.10$ to $1.40$. The plot in \fig{fig:phase_diagram} shows the structure of the trajectory with the lowest free energy at each point.

The phase diagram for our \gls{GAP} shown in \fig{fig:phase_diagram} reproduces very well the trends in the melting curve found experimentally~\cite{Morard2018}, including the raised melting temperature at $10$~GPa and the missing liquid phase at $100$~GPa up to $3000$~K. There is some disagreement between the melting temperature estimated using this method, which is located in the range $1500$--$1750$~K, and that estimated using the two-phase method in Sec.~\ref{sec:solid-liquid}, which is situated a bit below $1450$~K.
The \gls{fcc} phase is missing completely from our phase diagram, although it is found experimentally within a narrow band of temperatures, above $\sim 1180$~K and below $\sim 1670$~K (depending on the pressure).

At the highest pressures studied in this work, all the trajectories led to spontaneous nucleation of \gls{hcp} up to a temperature of at least $3000$~K (we did not check higher temperatures than this), also in agreement with the experimental data. In our phase diagram we also show the $c/a$-ratio, which is much lower than at ambient pressure. With increased temperature the spacing between the close-packed planes grows, which seems sensible. This leads us to believe that our \gls{GAP} could be suitable to study iron at the conditions of the Earth's core (exceeding $136$~GPa and approx. $4000$~K~\cite{Hirose2013}).
Under such conditions, the atomic volumes are in the range of the lowest volumes shown in \fig{fig:Stability}, corresponding to pressures in excess of $375$~GPa at a temperature of $6000$~K~\cite{Sun2022b}.

\begin{figure}[t]
\centering
    \includegraphics[width=1\linewidth,keepaspectratio]{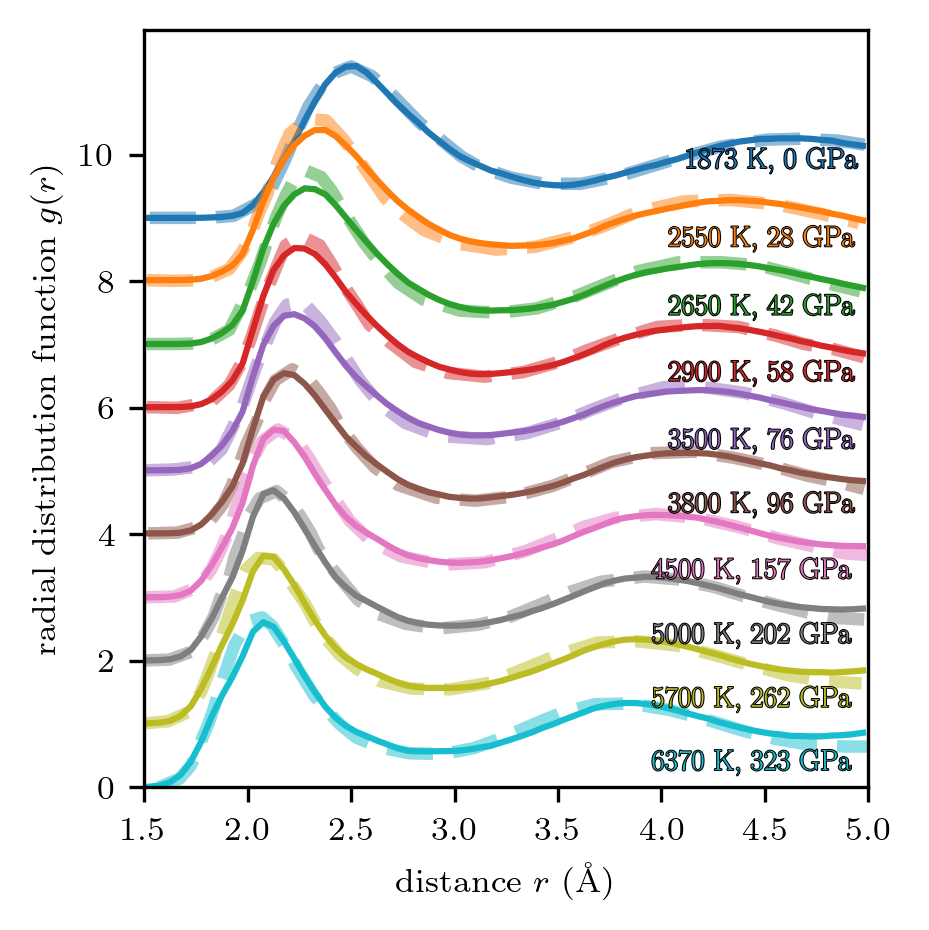}
    \caption{Liquid iron \glspl{RDF} at high ($p$, $T$) along the \gls{AIMD} melting line~\cite{Gonzalez2023} (dashed lines). Data computed using our \gls{GAP} overlaid (solid lines). Curves offset for clarity.}
    \label{fig:earths_core}
\end{figure}

To further elucidate the suitability of our \gls{GAP} to model the behavior of iron under extreme conditions, we computed the \glspl{RDF} of liquid iron at various ($p$, $T$) and compared them to reference data from \gls{AIMD} calculations~\cite{Gonzalez2023}, shown in \fig{fig:earths_core}.
Our \glspl{RDF} agree very well with the reference data, capturing the general shape of the curves as well as the shift of the first peak towards lower distances.

\section{Nanoparticles}\label{sec:nanoparticles}
\glspl{NP} were created in four distinct ways: 1) by condensation from a random starting atomic distribution, 2) using the \gls{GA} implementation by Weal \etal~\cite{Weal2021}, 3) with the Wulff method~\cite{Wulff1901} and 4) by annealing \glspl{NP} found with the other methods at elevated temperature.
For the first method, the atoms were randomly placed in a periodic box with double the atomic volume of the \gls{bcc} bulk. The atomic positions were then relaxed using the current version of the \gls{GAP} developed in this work using the algorithm detailed in the Atomic Simulation Recipes~\cite{Gjerding2021}. The suitable selection of the box volume ensured that the atoms coalesced into a single \gls{NP}.
The \gls{GA} was run with a population size of $100$ particles for $2500$ generations with $20$ offsprings per generation for each \gls{NP} size. The starting populations were created using the condensation method. For detailed \gls{GA} settings, see Sec.~VIII of the \gls{SM}~\cite{SM}. From all \glspl{NP} evaluated during a \gls{GA} run, the $1$st, $50$th, $100$th and $500$th lowest in energy (as per the \gls{GAP} developed in this work) were then calculated with \gls{DFT} to be used in the training of the next iteration of the \gls{GAP} potential. In total, four iterations of \gls{GA} were done to improve the performance of the potential for (increasingly) low-energy \glspl{NP}.
Crystalline \glspl{NP} were generated using the Wulff method~\cite{Wulff1901} as implemented in \gls{ASE}~\cite{Xiao2017}, using the surface energies we calculated using \gls{DFT}.
To augment the search space, \glspl{NP} found with the aforementioned methods were also annealed at $1200$~K for $20$~ps, quenched down to $300$~K over another $20$~ps and finally relaxed using gradient-descent minimization.

\begin{figure}[t]
%\centering
    \includegraphics[width=\linewidth,keepaspectratio]{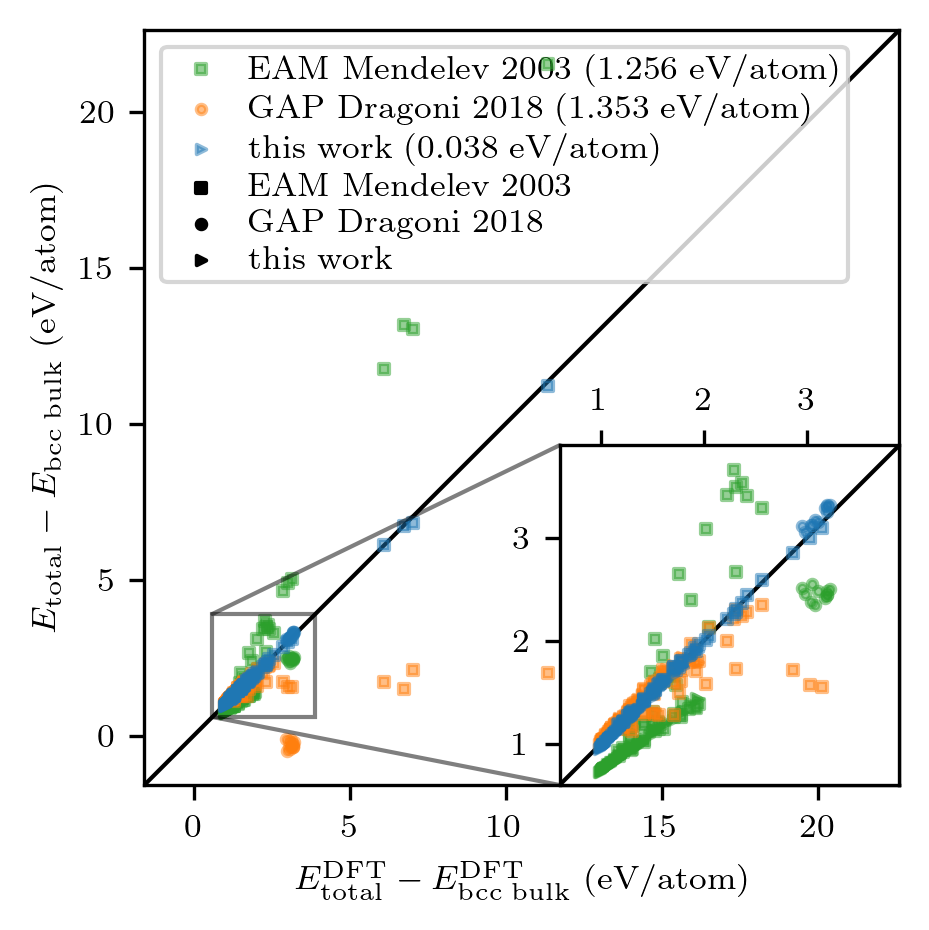}
    \caption{Energy for \glspl{NP} generated and computed with the GAP potential developed in this work, the \gls{GAP} potential by Dragoni \etal~\cite{Dragoni2018GAP} and the \gls{EAM} potential by Mendelev \etal~\cite{Mendelev2003} compared to \gls{DFT}. Shapes indicate the potential used to generate the \glspl{NP} and colors indicate the potential used to calculate the energies, respectively. The \glspl{NP} were generated using the random condensation method (for details, see text). None of the \glspl{NP} were used in the training of our GAP. The inset shows the data at lower energies more clearly.}
    \label{fig:NP_potentials}
\end{figure}

The region of configuration space corresponding to \glspl{NP} displays a rather complex \gls{PES}, due to the coexistence of diverse atomic motifs not encountered in the bulk: small surfaces, edges and vertices. To make this problem tractable, we used the iterative training approach~\cite{deringer_2017} combined with the \gls{GA} to incrementally improve the accuracy in this region of configuration space. Figure~\ref{fig:NP_potentials} shows the energies of the \glspl{NP} created by and calculated with our new \gls{GAP} and the two reference potentials, compared to the energies calculated using \gls{DFT} and referenced to the energy of the \gls{bcc} bulk.
The energies calculated with the two reference potentials differ significantly from the \gls{DFT} energy, while our \gls{GAP} predicts the energies with good accuracy, regardless of which potential was used to generate the \glspl{NP}. The \glspl{RMSE} are given in the legend.
Note how the Mendelev \gls{EAM} predicts too low energies for all the low-energy \glspl{NP} and the Dragoni \gls{GAP} predicts a number of \glspl{NP} to be lower in energy than the \gls{bcc} bulk material, again a sign of extrapolation outside of the training set, as seen for \gls{hcp} surfaces.
Note that none of the \glspl{NP} in \fig{fig:NP_potentials} belong to the \gls{GAP} training set, and thus this test gives a clear indication of the ability of our \gls{GAP} to accurately generate and predict iron \glspl{NP}, a particularly challenging modeling task.

In \fig{fig:NP_potentials} we show the accuracy of the energy predictions of our \gls{GAP} and the two reference potentials for a large number of \glspl{NP} over a wide range of stabilities. This is meant to ensure that our GAP can model small \glspl{NP} as well as ones that are far from the lowest energy for their size, as might be observed at the elevated temperatures of a catalytic process.
The inability of the \gls{bcc} reference potentials to correctly describe \glspl{NP} can be seen for example in the isolated group of points for the Dragoni \gls{GAP} at the bottom of the plot: all the \glspl{NP} in this group were created with the Dragoni \gls{GAP} and the performance on these is clearly different than on the \glspl{NP} created with the Mendelev \gls{EAM}.
In contrast to the two reference potentials, our \gls{GAP} performs very well on all the \glspl{NP} shown here, regardless of which potential they were created with or their relative stability. This can be measured in the \gls{RMSE} value which is about $30$ times lower than for the reference potentials. More importantly though, there is no region in the plot where our \gls{GAP} has substantial errors. This is particularly important for \gls{NP} modeling, when we use the \gls{GAP} to generate \glspl{NP} with a few hundred or thousands of atoms, which cannot be directly validated with \gls{DFT} due to CPU cost.

\begin{figure}[t]
\centering
    \includegraphics[width=\linewidth,keepaspectratio]{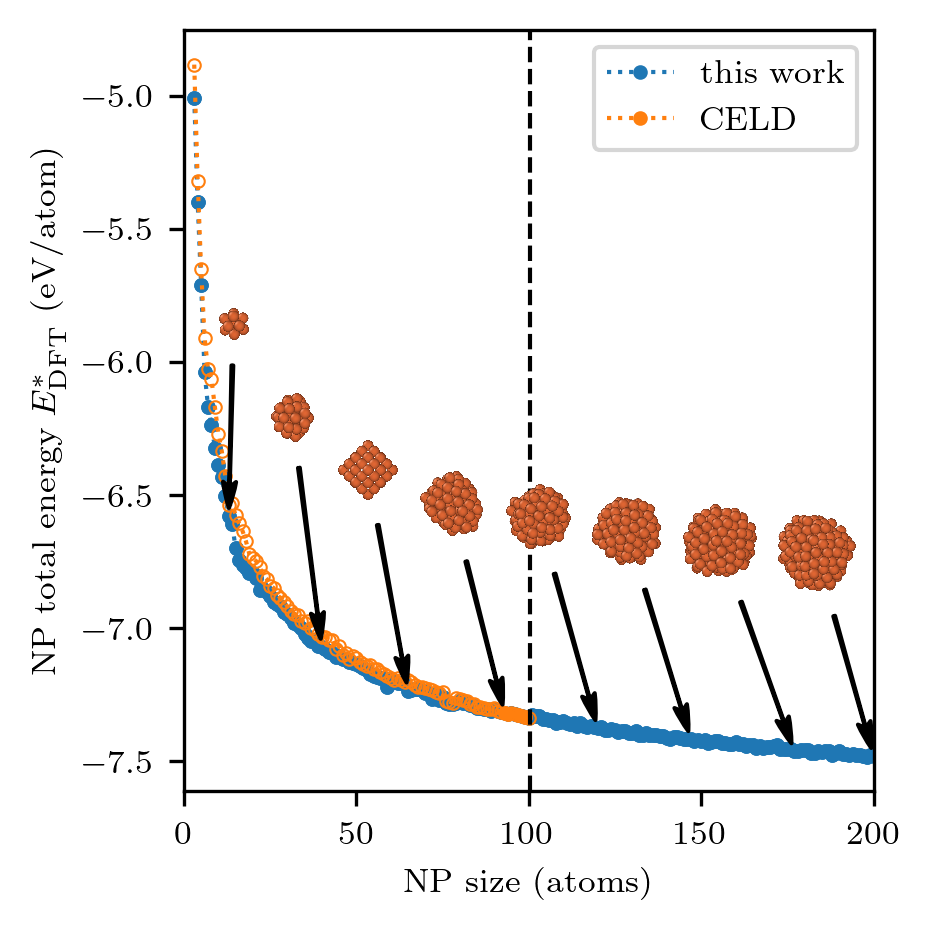}
    \caption{\gls{DFT} total energy convex hull of the \glspl{NP} from the Cambridge Energy Landscape Database~\cite{cambridge_energy_landscape} and the \glspl{NP} discovered in this work. Full circles indicate \glspl{NP} in the convex hull that were improved by this work ($90$ out of $98$ particles). Snapshots show a selection of \glspl{NP} along the convex hull.}
    \label{fig:convex_hull_DFT}
\end{figure}

% convex hull
In the search for stable \glspl{NP}, the common practice is to generate low-energy \glspl{NP} within a range of sizes which, for small \glspl{NP}, is measured in terms of the number of atoms. The energies are then used to construct a convex hull of \gls{NP} stability. Because of the high computational cost of \gls{DFT} calculations, the known convex hulls reach only up to a size of $30$ atoms at most~\cite{Bobadova-Parvanova2002,Kohler2005,Ma2007,Akturk2016}. More comprehensive convex hulls, up to a size of $100$ atoms, have been computed using the Finnis-Sinclair~\cite{FinnisSinclair,Sutton1990} \gls{EAM}~\cite{Elliott2009,Liu2016,cambridge_energy_landscape}. With our GAP, we have reconstructed the convex hull of the lowest-energy \glspl{NP} for each size up to a size of $200$ atoms, using the search methods detailed above. The \gls{GA} was only used up to a size of $100$ atoms, due to its comparatively high computational cost.
To validate the GAP results, the energies of the \glspl{NP} from the \gls{CELD}~\cite{cambridge_energy_landscape} as well as the \glspl{NP} in the convex hull found in this work were recalculated using \gls{DFT} up to a size of $200$ atoms. The total energies for both are shown in \fig{fig:convex_hull_DFT}. In the curve for the convex hull from this work, empty circles indicate \glspl{NP} that are higher in energy than the \gls{CELD} \gls{NP} of the same size and full circles such particles that are lower in energy.

\begin{figure}[t]
\centering
    \includegraphics[width=\linewidth,keepaspectratio]{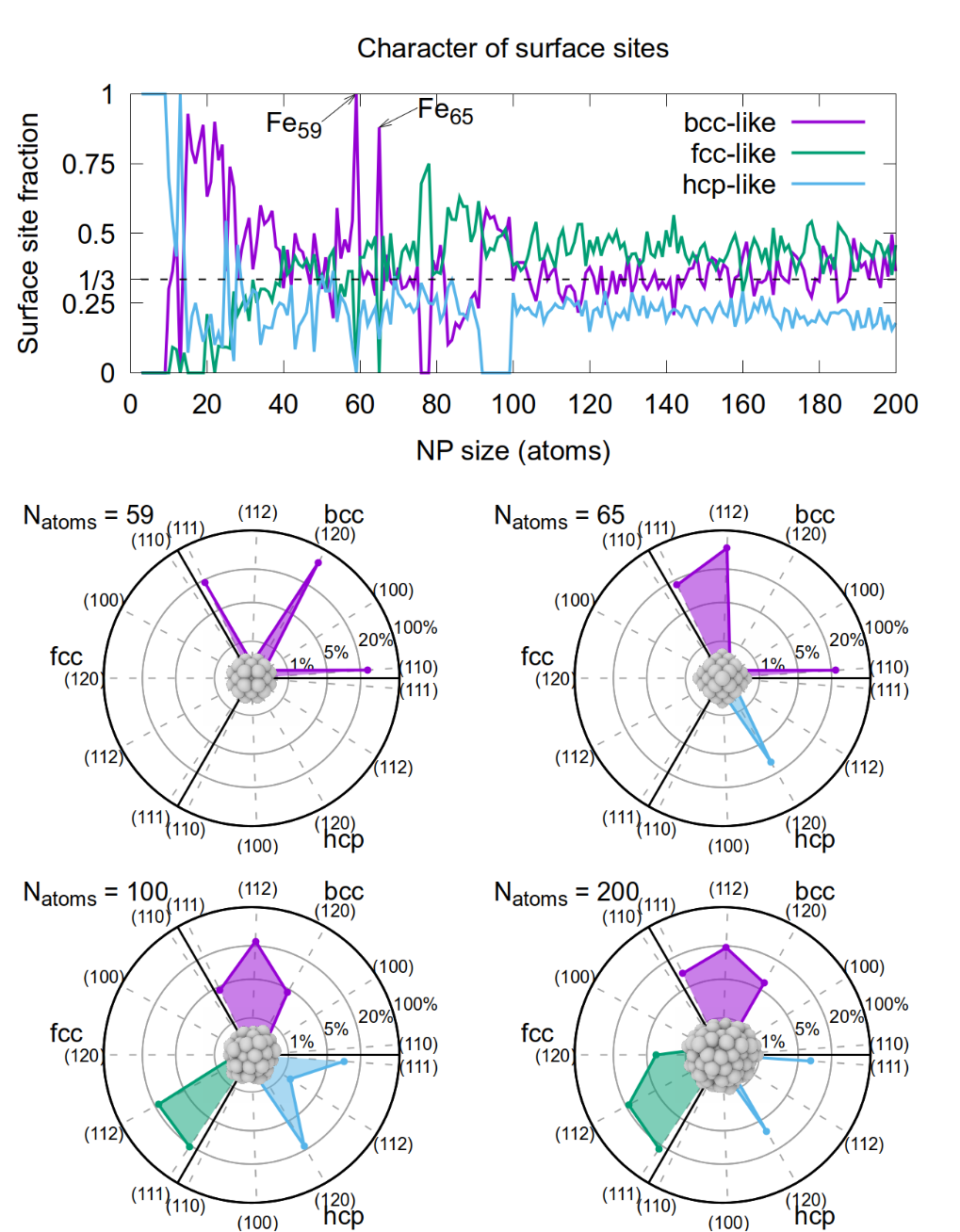}
    \caption{(Top) Fraction of surface sites that resemble [(100), (110), (120), (112) and (111)] surface motifs from pristine \gls{bcc}, \gls{fcc} and \gls{hcp} surfaces more than the others for all \glspl{NP} in our convex hull. (Bottom) Log-scaled fraction of surface sites that resemble the reference surfaces most for four example \glspl{NP}, two crystalline and two amorphous ones. A video with the panels corresponding to every \gls{NP} is available on Zenodo~\cite{zenodo_nps}.}
    \label{fig:character}
\end{figure}

\begin{figure}[t]
\centering
    \includegraphics[width=\linewidth,keepaspectratio]{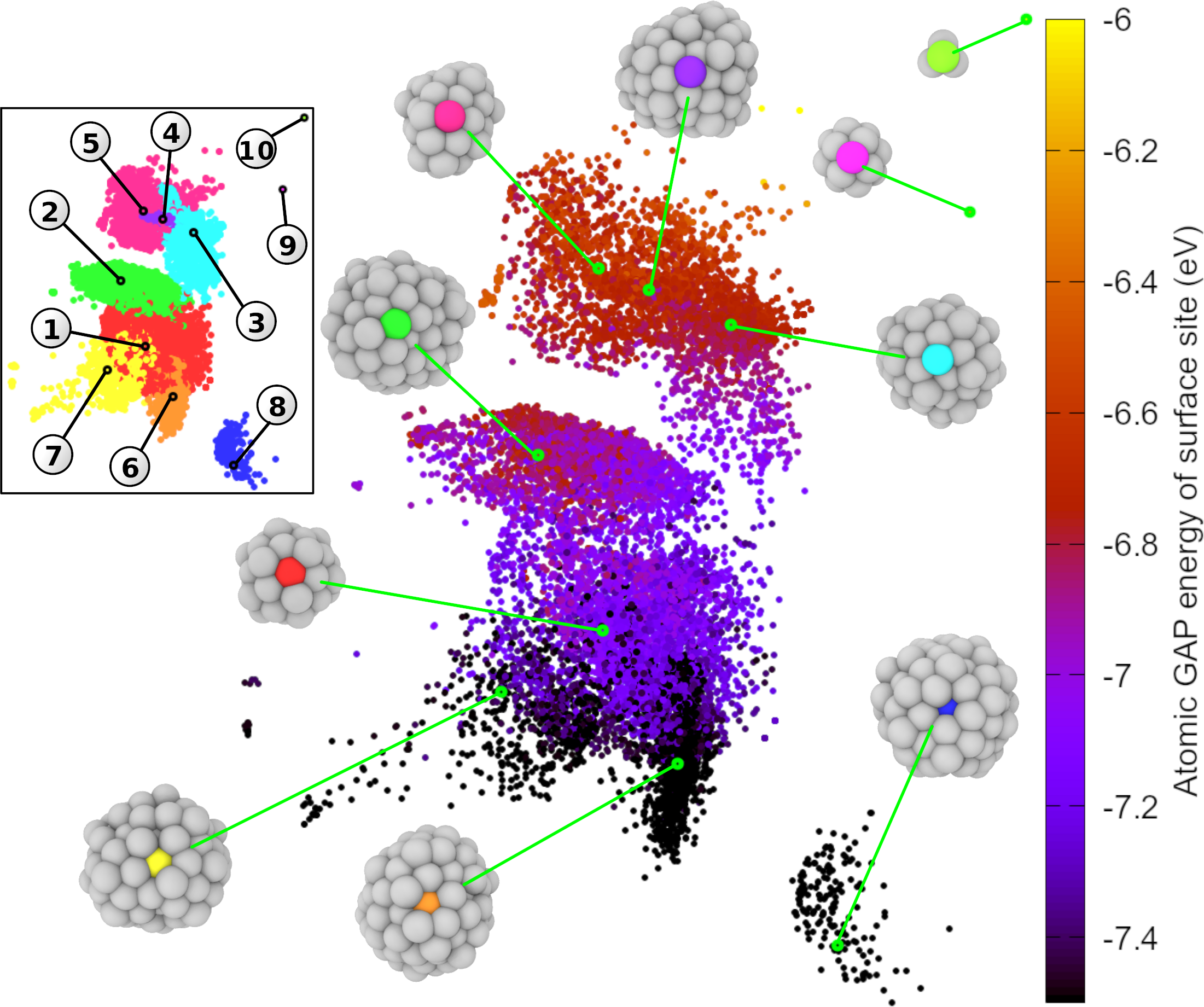}
    \caption{\texttt{cl-MDS} representation (low-dimensional embedding) of the surface sites on the \glspl{NP} in our convex hull clustered by $k$-medoids into ten characteristic motifs. The snapshots show the medoids representing the clusters. Encoded in the color is the atomic \gls{GAP} energy of each surface site. The inset shows the same map color coded according to the $k$-medoids clusters.}
    \label{fig:mds}
\end{figure}

\begin{figure*}[p]
\centering
    \includegraphics[width=\linewidth,keepaspectratio]{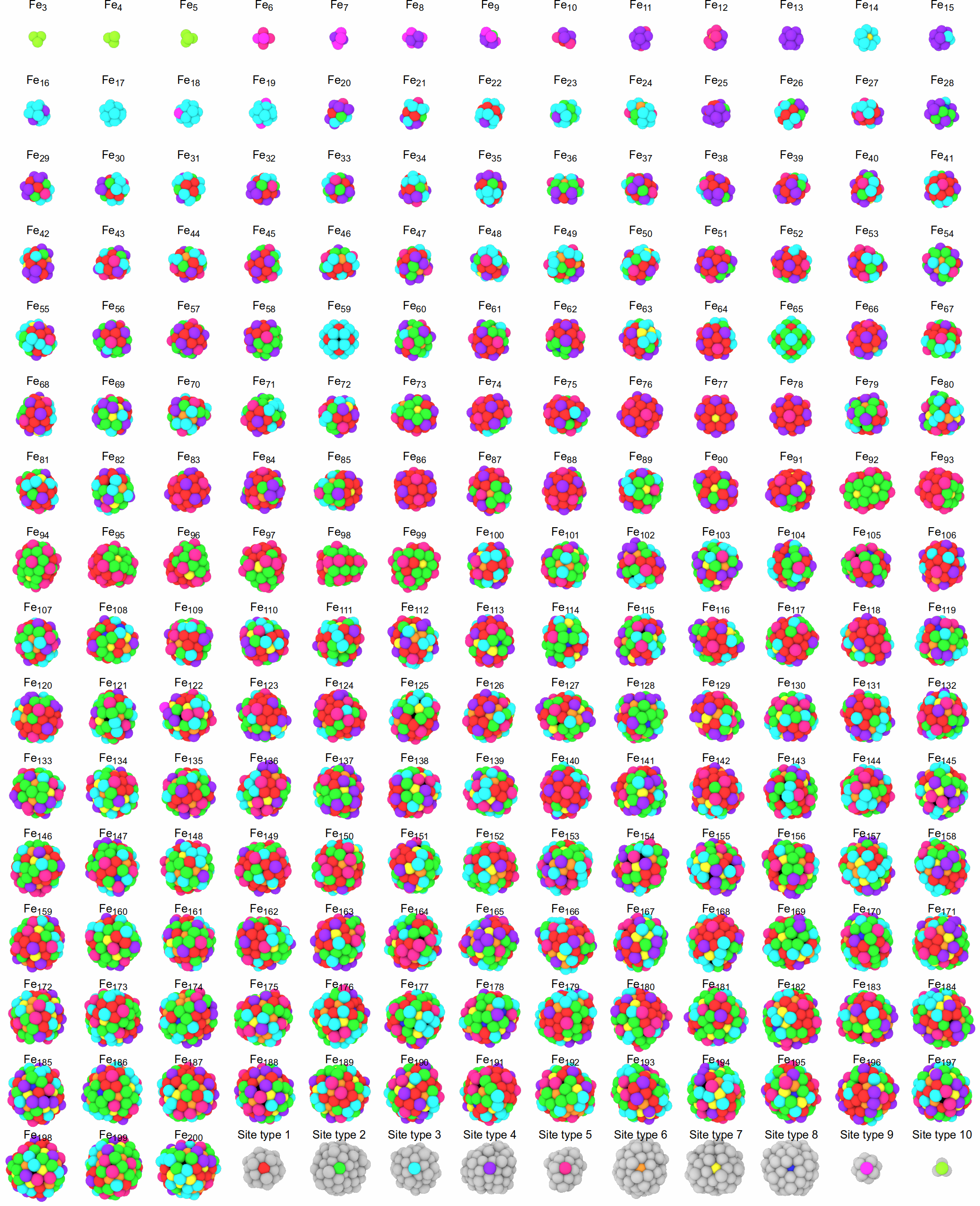}
    \caption{Snapshots of all \glspl{NP} from our convex hull, shown in \fig{fig:convex_hull_DFT}. Atoms are colored according to the ten most characteristic surface site motifs according to the $k$-medoids clustering shown in \fig{fig:mds}. The medoids for the ten motifs are shown as well.}
    \label{fig:all_nps}
\end{figure*}

In the size range from 3 up to 100 atoms, where \gls{CELD} data is available, $90$ out of $98$ \glspl{NP} in our convex hull are lower in energy than the \gls{CELD} ones. Most of these were found using the annealing method, but Fe$_{59}$ and Fe$_{65}$ were constructed with the Wulff method~\cite{Wulff1901} (subsequently relaxed using our \gls{GAP}).
From these, the two particles created using the Wulff method stand out from the curve with particularly low energy (e.g., see snapshot at $65$ atoms in \fig{fig:convex_hull_DFT}).
That we found more stable \glspl{NP} despite relying on a less sophisticated search method than Liu \etal~\cite{Liu2016}, in terms of the number of \glspl{NP} that could be sampled, can be attributed to the much higher accuracy of our potential compared to the Finnis-Sinclair \gls{EAM}. This is especially true for very small \glspl{NP}, where the error for the Finnis-Sinclair \gls{EAM} is the highest. (E.g., see the energies for the \glspl{CELD} convex hull shown in the SM~\cite{SM} Fig.~S7.)

Among the \gls{DFT} data reported in the literature, the structures of the \glspl{NP} are unavailable. We were thus unable to compute the total energies with our GAP for comparison. The published binding energies, on the other hand, strongly depend on the exchange-correlation functional used (e.g., BLYP by Ma \etal~\cite{Ma2007}, BLYP/SDD by Akt\"urk \etal~\cite{Akturk2016}) and are not easily comparable to our results. Hence, the convex hulls derived by Ma \etal{} and Akt\"urk \etal{} were omitted from \fig{fig:convex_hull_DFT}.

In the remainder of this section we try to uncover trends in the structure of these \glspl{NP}, in particular regarding surface features. We first analyze the similarity between \gls{NP} surface motifs and selected [(100), (110), (120), (112) and (111)] surface motifs present in pristine \gls{bcc}, \gls{fcc} and \gls{hcp} surfaces. As a first step, we identify surface atoms in the \glspl{NP} with a rolling-sphere algorithm as implemented in \texttt{ase\_tools}~\cite{ase_tools}. Then, the SOAP descriptors characterizing the environment of these surface atoms within a $4$~\AA{} sphere are computed, as well as the \gls{SOAP} descriptors of the atoms in the reference surfaces. Finally, we calculate the \gls{SOAP} kernels between \gls{NP} and surface descriptors, yielding a measure of similarity between $0$ and $1$. These kernels are used to rank the surface ``character'' of each \gls{NP} as a histogram counting the number of motifs of each type divided by the total number of surface sites on that \gls{NP}. The results of this analysis are given in \fig{fig:character} (top) for the overall \gls{bcc}/\gls{fcc}/\gls{hcp} character, whereas the bottom panel of the figure shows four examples further resolving the surface character for Fe$_{59}$, Fe$_{65}$, Fe$_{100}$ and Fe$_{200}$.

From the figure we infer that, except for very small \glspl{NP} with just a couple dozen atoms, the highly symmetric Fe$_{59}$ and Fe$_{65}$ corresponding to magic numbers, and the also highly symmetric Fe$_{76}$ and Fe$_{78}$, the motif distribution in these small \glspl{NP} is close to random, oscillating around an equal distribution of \gls{bcc}, \gls{fcc} and \gls{hcp} sites up to $N_\text{atoms}=100$. Beyond that, there are slightly more \gls{fcc} sites and slightly less \gls{hcp} sites. Given this degree of disorder, classifying surface sites in small Fe \glspl{NP} in terms of the crystalline surface motifs is not very useful. Instead, we resort to a motif classification scheme that draws the classes directly from the database of structures. We do this using $k$-medoids, a data-clustering technique that separates data points into classes (``clusters'') according to their similarity~\cite{bauckhage_2015}. The most representative data point in each cluster is called a medoid. In our case, a collection of medoids provides a catalogue of representative motifs found in our database~\cite{caro_2018c}. We used the \texttt{fast-kmedoids} library for the $k$-medoids computation~\cite{fast_kmedoids} and \texttt{cl-MDS} to graphically represent the clustering via low-dimensional embedding~\cite{hernandez-leon_2022}. The results are shown in \fig{fig:mds}.

In the figure we classified the surface sites of all the lowest-energy \glspl{NP} for a given number of atoms, from $3$ to $200$ atoms, into 10 data clusters. The size of the clusters decreases with their number, i.e., cluster 1 represents the most common surface motifs and 10 the least common. The color encodes the local GAP energy of the surface atoms in the big map, and the cluster index in the figure inset, for better reference. We observe ``burried'' (almost sub-surface) motifs to be the most stable (clusters 6, 7 and 8, $\approx -7.5$~eV/atom). Then, motifs with a six-fold surface coordination (i.e., they appear to be at the center of a hexagon on the surface) but embedded within the \gls{NP} facet are contained in clusters 1 ($\approx -7.2$~eV/atom) and 2 ($\approx -6.9$~eV/atom). Six-fold coordinated motifs that are raised further from the surrounding atoms are higher in energy, and contained within cluster 5 ($\approx -6.5$~eV/atom). Five-fold coordinated motifs are similar in energy to the latter, with the central atom similarly raised, and belong to clusters 3 and 4. Finally, clusters 9 and 10 contain just a handful of motifs found in the extremely small \glspl{NP}.

The atomic GAP energy of bulk \gls{bcc} iron is $\approx -8.25$~eV/atom. The energy difference between this bulk value and the GAP energy of a less stable motif is directly related to the cohesive energy that could be gained by, e.g., increasing the coordination of the less stable motif. Therefore, we expect the less stable motifs to readily passivate while in contact with a surrounding medium, e.g., by forming strong bonds with and effectively capturing hydrogen atoms. Opposed to this, some of the motifs (especially those in clusters 6, 7 and 8) may be too stable to interact with adsorbants via covalent interactions. The sites with intermediate atomic GAP energies might be the most interesting from the catalytic point of view, e.g., because of their potential to adsorb or desorb reactants as a function of applied external bias. We will explore the precise relationship between adsorption energy of typical adsorbants and atomic GAP energies in subsequent work on Fe \gls{NP} reactivity.

Finally, a gallery of all the \glspl{NP} in our convex hull database is given in \fig{fig:all_nps}, with each surface site colored according to the data cluster to which it belongs (the reference motifs are also shown in the figure). We can easily observe that, except for the highly symmetric \glspl{NP} at very small size and magic numbers $59$ and $65$, as well as the stability island between $76$ and $78$ atoms, the distribution of surface motifs is highly irregular. That is, there is no obvious facet formation in these \glspl{NP}. Tests that we carried out for a significantly larger \gls{NP} with a few thousand atoms, generated using the condensation method, also showed lack of significant facet formation. This contrasts with the very clear facet formation in other metal \glspl{NP}, for instance (111) facets in Pt \glspl{NP} as we have recently observed using very similar methodology~\cite{kloppenburg_2023}. A possible explanation for this is that structural disorder in iron \glspl{NP} is driven by the interplay between the formation of the stable \gls{fcc} surface facets versus the formation of the stable \gls{bcc} bulk motifs. Since the bulk motif will nucleate facet formation with its same crystal structure and vice versa, this may lead to a non-trivial dynamics which in turn results in highly disordered \glspl{NP}. Indeed, it has been shown experimentally that nanostructured Fe, e.g., Fe thin films on a substrate, can be grown in the \gls{fcc} structure even at room temperature~\cite{li_1994,keavney_1995}.

\section{Code and data availability}
The GAP is available for free on Zenodo~\cite{zenodo_gap} and can be used with QUIP/GAP, LAMMPS via the QUIP interface, ASE via Quippy, and TurboGAP.
Incidentally, we note an improvement in computational efficiency of our \gls{GAP} over the previous state-of-the-art Dragoni \gls{GAP} by a factor of approx. $4$. This speedup can be attributed mostly to the use of SOAP descriptor compression~\cite{darby_2022,darby_2022b} in our \gls{GAP}, as available from the \texttt{soap\_turbo} descriptor~\cite{soap_turbo}. When used with the TurboGAP \gls{MD} engine~\cite{Caro2019}, better speedups can usually be achieved.

To facilitate further work in this area, we have made the structures of the \glspl{NP} derived in this work available to the community. A full database is available for download on Zenodo~\cite{zenodo_nps}, including the energies computed with the reference potentials and the Finnis-Sinclair \gls{EAM}.

\section{Summary}
In summary, we created a generally applicable \gls{GAP} \gls{ML} potential for the iron system which is stable in the whole configuration space and performs well for a wide range of applications, from bulk to nanostructured iron, from ambient conditions to those at the Earth's core. The accurate description of nanoparticles at elevated temperatures is particularly useful for the simulation of catalytic processes, which often occur at those temperatures.
A straight-forward approach for such studies would be to combine our \gls{GAP} with another \gls{GAP} for the other involved species and extend the training databases with mixed structures, followed by some iterative training.
While our \gls{GAP} cannot beat previously existing specialized potentials in every case, it can be used reliably for most problems, including the study of systems where two or more Fe phases coexist. We found it to be the most accurate for \glspl{NP} from among the potentials considered. We have derived a series of low-energy Fe \glspl{NP} and made these structures available for further use.
The \gls{GAP} potential itself, which in addition to accuracy also achieves a sizeable speedup over the previous state-of-the-art potential, has also been made freely available. We hope that this will enable and stimulate further work in this field, in particular with regard to catalytic applications on low-dimensional iron structures.

Some limitations remain from the implicit treatment of the magnetic states, especially regarding surface energies and elastic constants. These could be addressed by training a \gls{GAP} including an explicit description of atomic magnetic moments. To this end, the necessary methodology and infrastructure to treat magnetism explicitly within the \gls{GAP} framework needs to be developed.

\begin{acknowledgments}
The authors are grateful to the Academy of Finland for financial support under projects \#321713 (R.~J. \& M.~A.~C.) and \#330488 (M.~A.~C.), and CSC -- IT Center for Science as well as Aalto University's Science-IT Project for computational resources.
\end{acknowledgments}

\bibliography{references} % for overleaf

\end{document}

\title{Supplemental material: Searching for iron nanoparticles \\ with a general-purpose Gaussian approximation potential}

\author{Richard Jana}
\email{richard.jana@aalto.fi}
\affiliation{Department of Chemistry and Materials Science, Aalto University, 02150, Espoo, Finland}

\author{Miguel A. Caro}
\affiliation{Department of Chemistry and Materials Science, Aalto University, 02150, Espoo, Finland}

\date{\today}

\maketitle
\beginsupplement

This supplemental material contains miscellaneous technical detail pertaining our manuscript, such as input parameters for reproducing our calculations and complementary figures.

\section{GAP fit command}\label{subsec:gap_fit}
The following command for gap\_fit was used to train our \gls{GAP}:

\begin{widetext}\begin{lstlisting}
gap_fit atoms_filename=train_all_pbe_tagged.xyz core_param_file=core_pot.xml \
    core_ip_args={IP Glue} \
    gap={ distance_2b Z1=26 Z2=26 cutoff=5.0 n_sparse=40 covariance_type=ard_se \
                delta=1. theta_uniform=0.5 sparse_method=uniform add_species=F: \
          angle_3b Z_center=26 Z1=26 Z2=26 cutoff=3.0 n_sparse=200 covariance_type=pp \
                delta=0.01 theta_uniform=4.0 sparse_method=uniform add_species=F: \
          soap_turbo l_max=8 alpha_max={{8}} atom_sigma_r={{0.4}} atom_sigma_t={{0.4}} \
                atom_sigma_r_scaling={{0.}} atom_sigma_t_scaling={{0.}} zeta=4 \
                rcut_hard=5.0 rcut_soft=4.5 basis="poly3gauss" scaling_mode="polynomial" \
                amplitude_scaling={{1.0}} n_species=1 species_Z={26} central_index=1 \
                radial_enhancement={{1}} compress_file="compress.dat" \
                central_weight={{1.0}} config_type_n_sparse={...} delta=0.1 f0=0.0 \
                covariance_type=dot_product sparse_method=cur_points add_species=F } \
        default_sigma={0.0002 0.02 0.02 0.02} energy_parameter_name=free_energy \
        force_parameter_name=NULL virial_parameter_name=NULL sparse_jitter=1.0e-8 \
        e0=-3.406721695 do_copy_at_file=F sparse_separate_file=T gp_file=iron.xml
\end{lstlisting}\end{widetext}

where the number of sparse configurations per config type is given in \tab{tab_s:db_composition}. Trivial compression~\cite{Caro2019} was used with a compression file. $e0$ was chosen as half the energy of a dimer at the cutoff distance $5$~\AA{}.

\begin{table*}[htb]
    \centering
    \caption{Number of sparse configurations and sigma parameters for each configuration type used in the training of the \gls{GAP}.}
    \begin{tabular}{ l | c | c | c | c }
        \hline
        config type & config\_type\_n\_sparse & number of configs in DB & $\sigma$ energy (eV/atom) & $\sigma$ virial (eV/atom) \\ 
        \hline
        dimer & $44$ & $44$ & $0.001$ & $0.1$ \\ %54
        trimer & $232$ & $232$ & $0.001$ & $0.1$ \\ %300
        bcc cell & $275$ & $4196$ & $0.001$ & $0.1$ \\
        fcc cell & $275$ & $6165$ & $0.001$ & $0.1$ \\
        hcp cell & $275$ & $1040$ & $0.001$ & $0.1$ \\
        bcc FM elastic & $800$ & $1918$ & $10^{-5}$ & $0.001$ \\
        fcc AFM elastic & $322$ & $322$ & $10^{-5}$ & $0.001$ \\ %800
        hcp NM elastic & $399$ & $399$ & $10^{-5}$ & $0.001$ \\ %800
        bcc rattle & $100$ & $2200$ & $10^{-5}$ & $0.001$ \\
        fcc rattle & $100$ & $409$ & $10^{-5}$ & $0.001$ \\
        hcp rattle & $100$ & $175$ & $10^{-5}$ & $0.001$ \\
        nanoparticles & $500$ & $1664$ & $0.001$ & $0.1$ \\
        nucleation cluster & $25$ & $331$ & $0.001$ & $0.1$ \\
        bcc FM surface & $127$ & $127$ & $0.001$ & $0.1$ \\ %500
        fcc FMLS surface & $216$ & $216$ & $0.001$ & $0.1$ \\ %500
        hcp NM surface & $72$ & $72$ & $0.001$ & $0.1$ \\ %500
        thin slab & $38$ & $38$ & $0.025$ & $2.5$ \\ %
        transition & $25$ & $42$ & $0.001$ & $0.1$ \\
        vacancy interstitial & $25$ & $26$ & $$ & $$ \\
        vacancy migration & $25$ & $42$ & $$ & $$ \\
        melt & $75$ & $292$ & $0.001$ & $0.1$ \\
        simple cubic & $50$ & $596$ & $0.025$ & $2.5$ \\
        diamond & $75$ & $475$ & $0.025$ & $2.5$ \\
        \hline
    \end{tabular}
\label{tab_s:db_composition}
\end{table*}

Furthermore, the database has been pre-processed by removing all but $25\%$ of the forces and scaling the the rest by $0.05$, but to a minimum of $0.1$~eV/\AA{}. Energy and virial regularization is done on a per-structure basis with the parameters given in \tab{tab_s:db_composition}

Dimer and trimer structures in the database were limited to energies of $100$ and $250$~eV/atom, respectively, to avoid impacting the accuracy of the potential in more stable regions of phase space. Thus, the inter-atomic distance of the closest dimer in the database is $0.7$~\AA{}.

\section{Stability plot}
\begin{figure}
    \includegraphics[width=\linewidth,keepaspectratio]{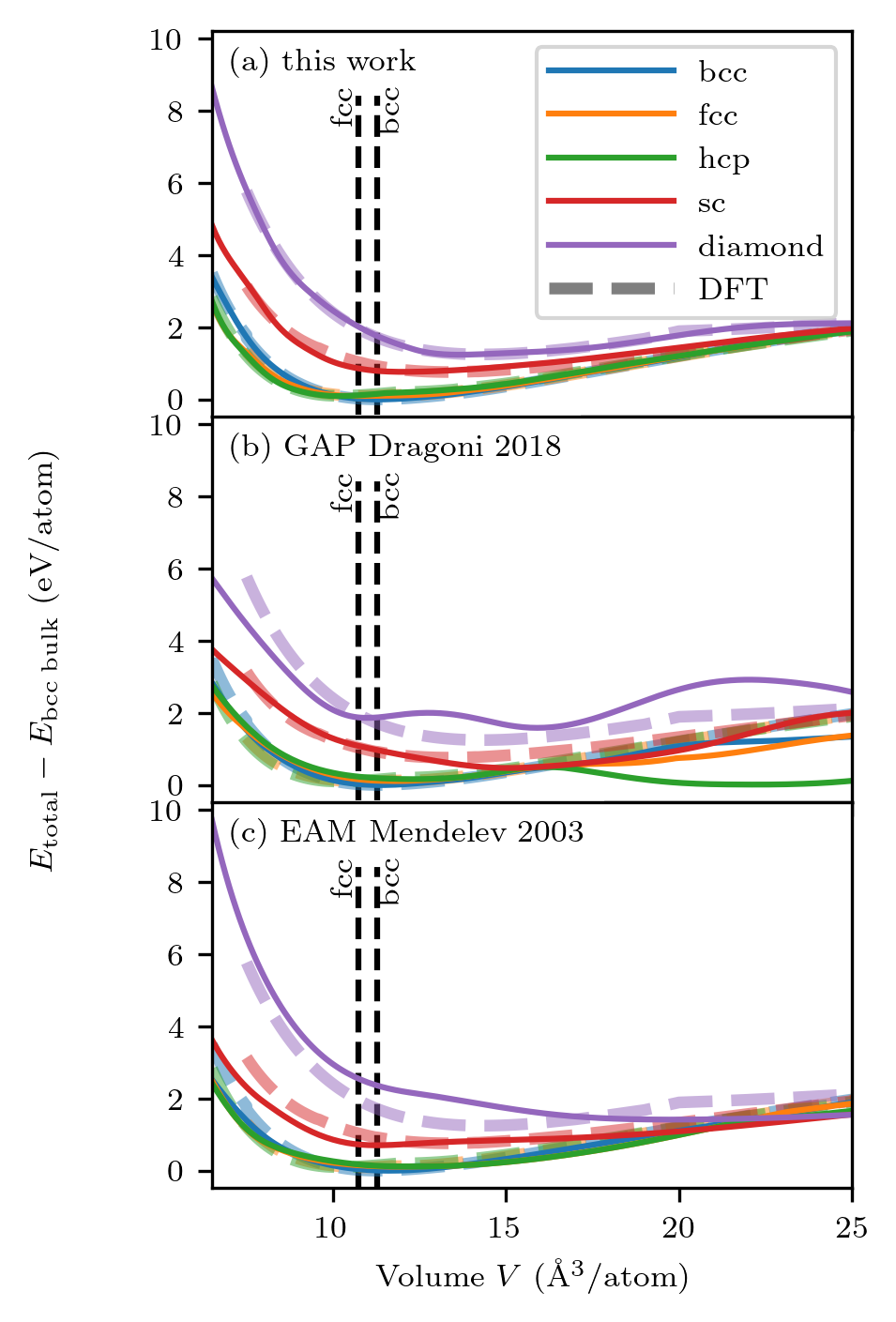} \\
    \caption{Plots corresponding to Fig.~2  over a larger range of atomic volumes.}
    \label{fig_s:Stability}
\end{figure}

\begin{figure}
    \includegraphics[width=\linewidth,keepaspectratio]{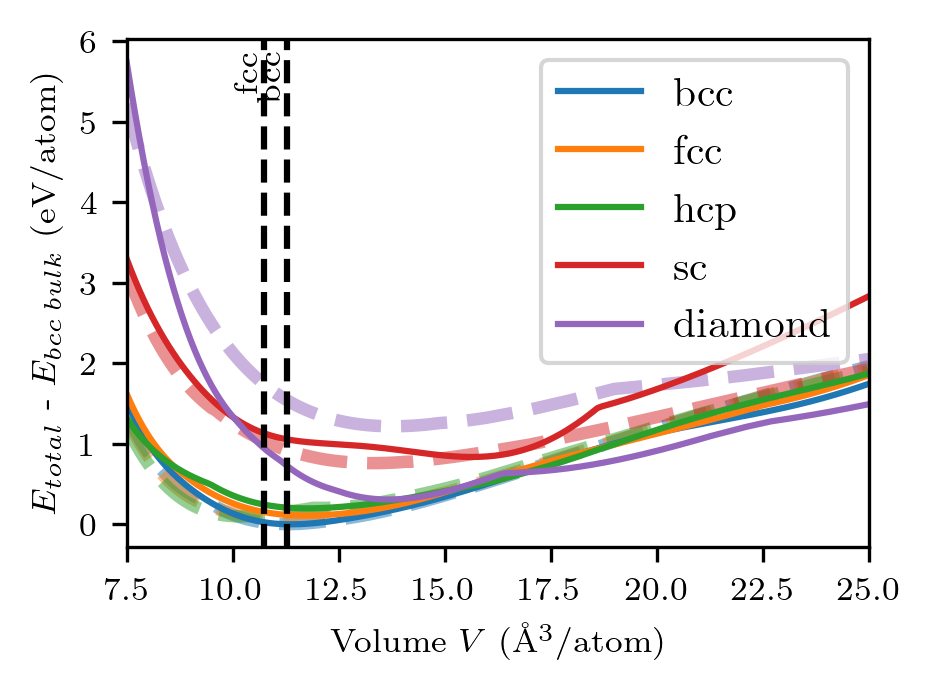} \\
    \caption{Stability of different crystal phases for the fracture \gls{GAP}~\cite{Zhang2022}. The \gls{hcp} minimum around $22$~\AA{}$^3$/atom has been fixed.}
    \label{fig_s:Stability_zhang22}
\end{figure}

\fig{fig_s:Stability} shows an extension of Fig.~2 in the main text to larger atomic volumes. For the Dragoni \gls{GAP}, this reveals a minimum in the \gls{hcp} curve around $22$~\AA{}$^3$/atom which is almost as low in energy as the global \gls{bcc} minimum. This has been fixed in the fracture \gls{GAP} by Zhang \etal{}~\cite{Zhang2022}, based on the Dragoni \gls{GAP} (see \fig{fig_s:Stability_zhang22}).
For both potentials the predictions for diamond and \gls{sc} structures are not accurate, but high enough in energy to be unproblematic.

\section{Cell parameters}
The cell parameters given in \tab{tab_s:cell_params} were obtained by relaxing the \gls{DFT} minimum structures with the respective potentials. Hence, other local minima in the cell parameter space might exist.

\begin{table*}[htb]
    \centering
    \caption{Cell parameters for the different magnetic configurations of \gls{fcc}, relaxed with \gls{DFT} and the three potentials. Values missing from the table indicate that no stable structure was found close to the \gls{DFT} cell parameters.}
    \begin{tabular}{ l | c c c c }
        \hline
         & DFT & this work & Dragoni GAP~\cite{Dragoni2018GAP} & Mendelev EAM~\cite{Mendelev2003} \\
        \hline
        bcc FM $a$ (\AA{}) & $2.825$ & $2.829$ & $2.834$ & $2.855$ \\
        \hline
        fcc AFM $a$ (\AA{}) & $3.418$ & $3.418$ & $3.416$ &  \\
        fcc AFM $c$ (\AA{}) & $3.674$ & $3.684$ & $4.042$ &  \\
        fcc FMLS $a$ (\AA{}) & $3.474$ & $3.484$ &  &  \\
        fcc FMHS $a$ (\AA{}) & $3.634$ &  & $3.611$ & $3.658$ \\
        \hline
        hcp NM $a$ (\AA{}) & $2.455$ & $2.457$ & $2.563$ & $2.621$ \\
        hcp NM $c$ (\AA{}) & $3.884$ & $3.886$ & $4.332$ & $4.146$ \\
        hcp NM $c \slash a$ & $1.582$ & $1.582$ & $1.690$ & $1.582$ \\
        hcp NM $c \slash a$ ($\sqrt{8 \slash 3}$) & $0.969$ & $0.969$ & $1.035$ & $0.969$ \\
        \hline
    \end{tabular}
\label{tab_s:cell_params}
\end{table*}

\section{Surface energies}
\begin{figure}
\centering
    \includegraphics[width=0.66\linewidth,keepaspectratio]{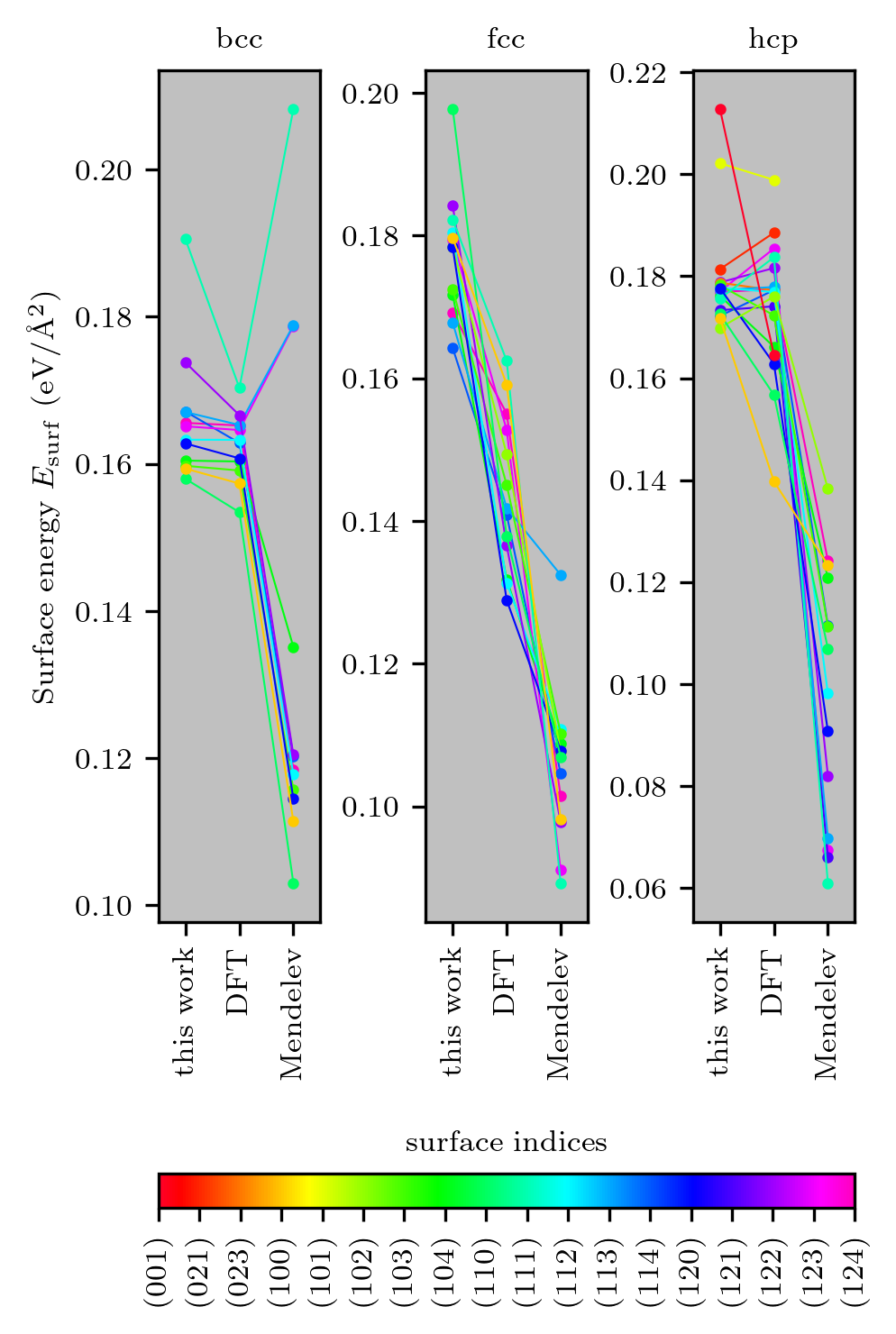}
    \caption{Surface energies of a wide range of surface indices for the \gls{EAM} by Mendelev \etal{} and our \gls{GAP}, compared with the \gls{DFT} values.}
    \label{fig_s:surface_energies_Mendelev}
\end{figure}

Fig.~\ref{fig_s:surface_energies_Mendelev} shows a plot equivalent to Fig.~6 in the main text. Instead of the Dragoni \gls{GAP}, this plot shows the Mendelev \gls{EAM} as reference potential, compared with our \gls{GAP} and the \gls{DFT} values.

\section{Thermal expansion}
\begin{figure}
\centering
    \includegraphics[width=\linewidth,keepaspectratio]{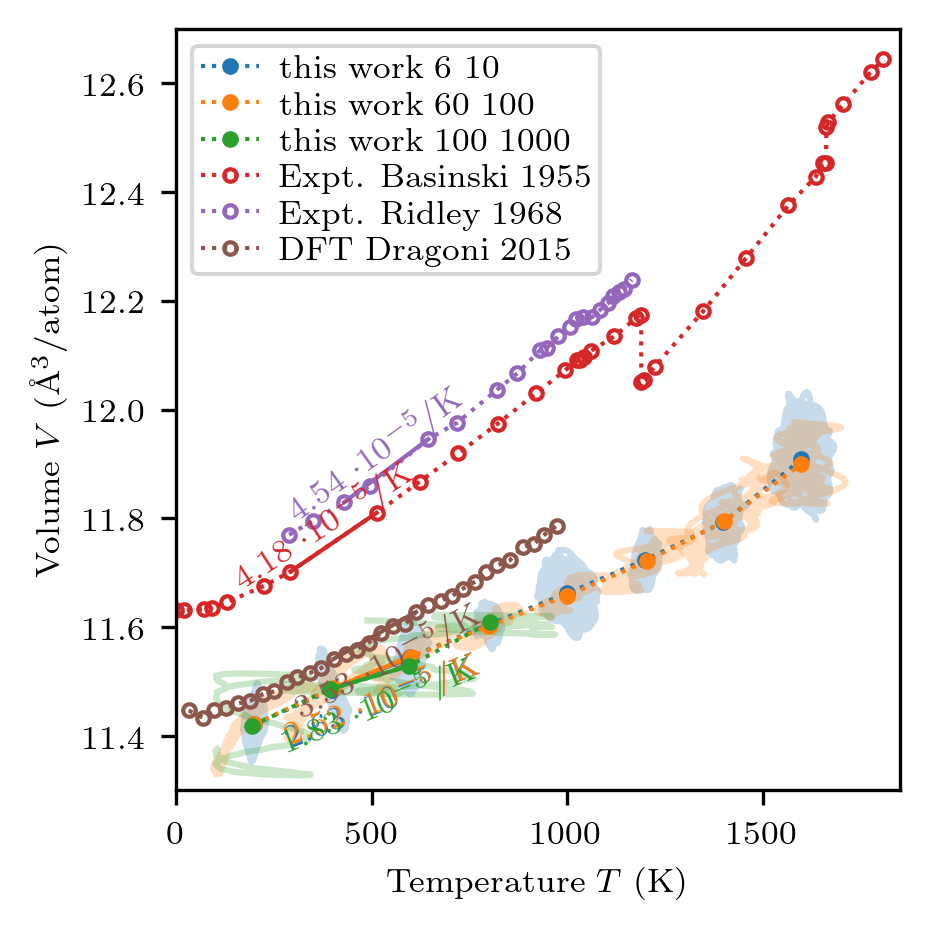}
    \caption{The thermal expansion plot, with two additional curves for our GAP, computed with longer characteristic timescales. The curves for the Mendelev \gls{EAM} and Dragoni \gls{GAP} were remove for clarity.}
    \label{fig_s:thermal_expansion_params}
\end{figure}

During equilibration and averaging of the thermal expansion calculations (see Fig.~7 in the main text) the target temperature and pressure were maintained using Nos\'e-Hoover and Parrinello-Rahman dynamics~\cite{Melchionna1993,Melchionna2000,Holian1990,DiTolla1993}, with the barostat bulk modulus set to $100$~GPa and the thermostat and barostat characteristic timescales set to $6$~fs and $10$~fs, respectively.
The short timescales were chosen for computational efficiency, but do not affect the results as compared to longer timescales. This can be seen in \fig{fig_s:thermal_expansion_params}, which compares sets of different characteristic timescales for thermostat and barostat.
The parameter sets (6, 10) and (60, 100) were run for the times stated in the main text. (100, 1000) was equilibrated for $10$~ps ($10$ times longer) and averaged over $2$~ps at each temperature.
While the oscillations in temperature and pressure increase with larger parameters, clearly, the different characteristic timescales do not impact the average values at any temperature. 

\section{Melting temperature}
\begin{figure}
    \includegraphics[width=\linewidth,keepaspectratio]{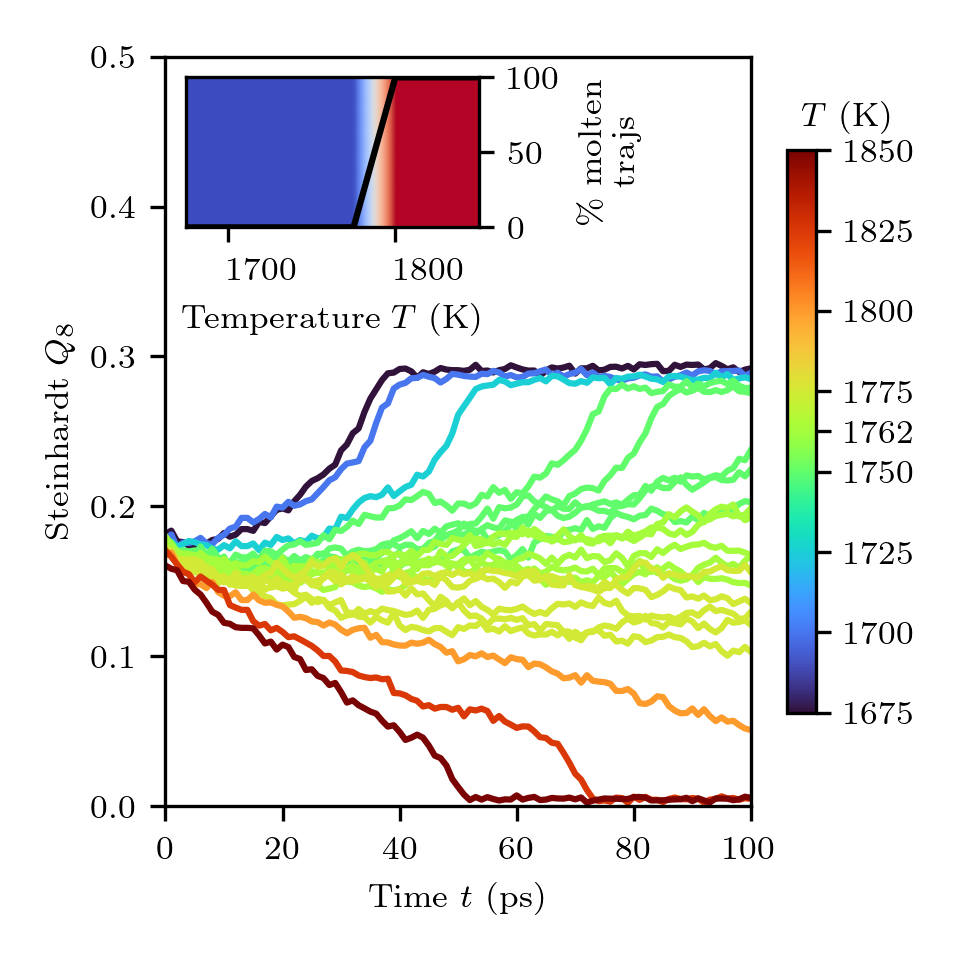} \\
    \caption{Two-phase trajectories to determine the melting temperature for the \gls{EAM} potential by Mendelev \etal{}~\cite{Mendelev2003}.}
    \label{fig_s:Tmelt_references}
\end{figure}

Fig.~\ref{fig_s:Tmelt_references} shows the results for the Mendelev \gls{EAM}~\cite{Mendelev2003} potential. Each line represents a separate \gls{MD} run at the temperature indicated by the color of the line.
Multiple calculations were run at each temperature with different random seeds to account for random thermal fluctuations ($5$ trajectories in the temperature range from $1750$ to $1775$~K and $1$ at all other temperatures shown).

\section{Phase diagram}
\begin{figure}
    {(a) reference structures} \\
    \includegraphics[width=\linewidth,keepaspectratio]{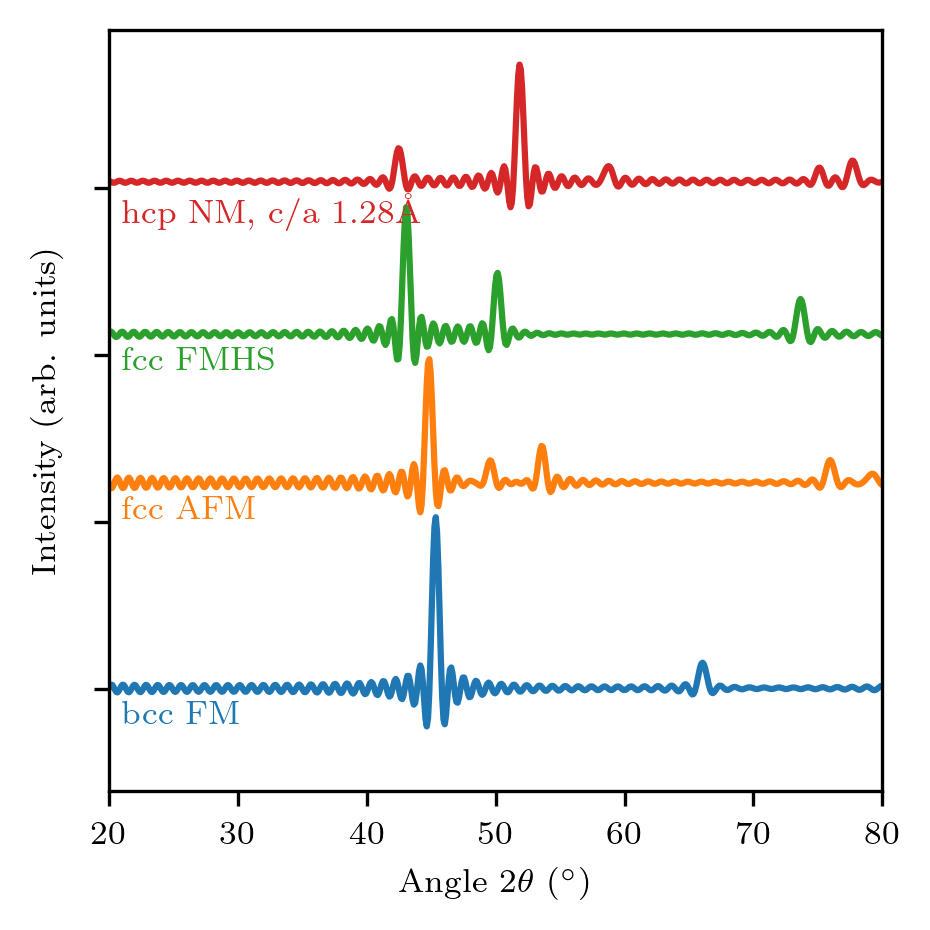} \\
    {(b) $p = 1$~bar} \\
    \includegraphics[width=\linewidth,keepaspectratio]{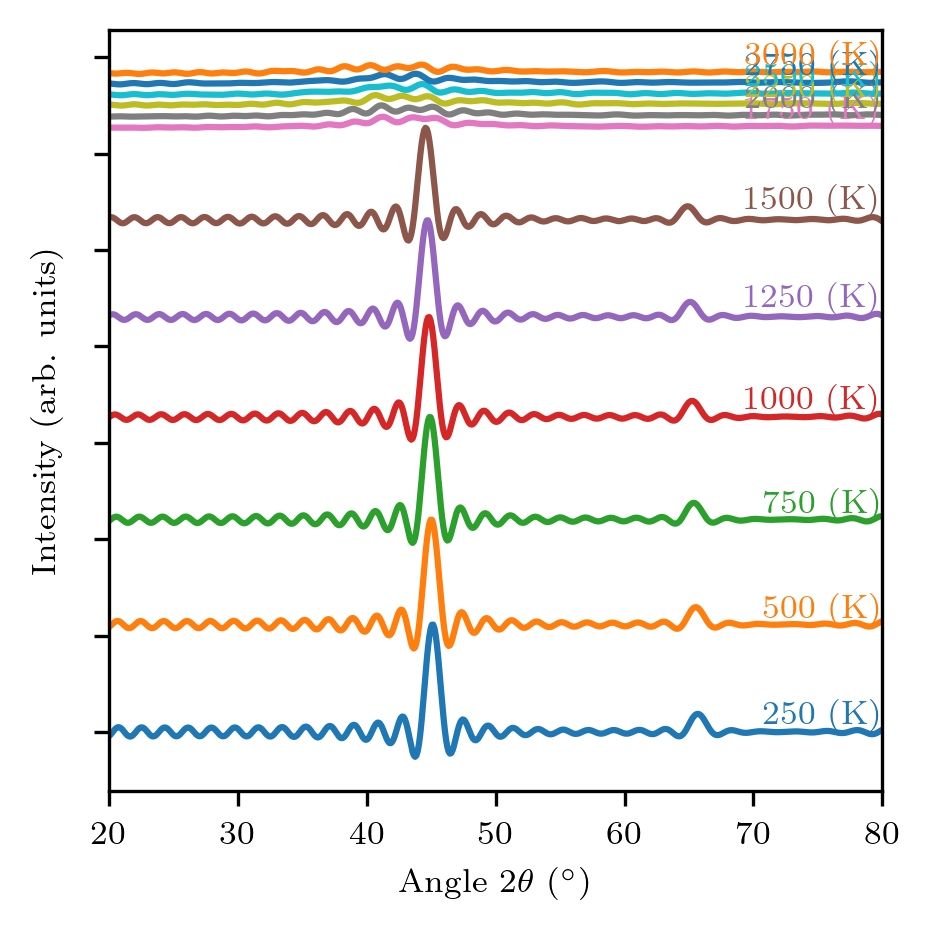}
    \caption{Example \gls{XRD} spectra for (a) the reference structures of different crystalline phases and (b) spectra at $p = 1$~bar over the whole temperature range.}
    \label{fig_s:ref_spectra}
\end{figure}

To label the phase diagram for our \gls{GAP}, shown in Fig.~9 in the main text, the \gls{XRD} spectra of the lowest-energy structure at each (p, T) were computed~\cite{ref_debye} and compared to the reference spectra shown in \fig{fig_s:ref_spectra} (a), representing perfect crystalline structures.
\fig{fig_s:ref_spectra} (b) shows the spectra for a slice through the phase diagram at $p = 1$~bar. Up to $1500$~K, all structures are \gls{bcc}, and above liquid.

\section{Solid-liquid interfaces}
\begin{figure}
    \centering
    \includegraphics[width=\linewidth,keepaspectratio]{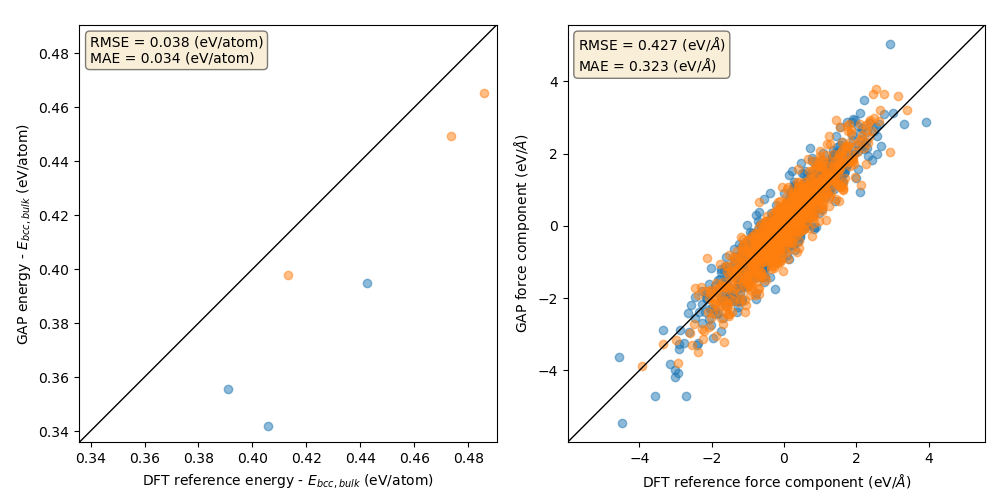} \\
    \caption{Energy of force errors of our \gls{GAP} compared to \gls{DFT} for solid-liquid interfaces created with our \gls{GAP} (blue) and the Mendelev \gls{EAM} (orange).}
    \label{fig:interface_energies}
\end{figure}

We employ two different methods to study the melting transition and phase diagram, the 2-phase coexistence approach and the free-energy calculations, relying on different approximations. As we found some level of disagreement between the two methods, we tested the performance of our \gls{GAP} on liquid-solid interfaces not included in the training set. Taking into account the number of different crystalline, surface  and amorphous structures as well as melt in our training data set, we would expect reasonable performance on liquid-solid interfaces.
 However, to demonstrate this, we prepared small systems ($108$ atoms) in a way similar to the 2-phase method we used in the main manuscript and tested our \gls{GAP} on those, compared to \gls{DFT}. To avoid bias, such interfaces were created with our \gls{GAP} and with the Mendelev \gls{EAM}.
\fig{fig:interface_energies} shows the energies and forces computed with our \gls{GAP} compared to \gls{DFT}.

Our \gls{GAP} underestimates the energy of these interfaces just slightly, with an \gls{RMSE} similar to that of configuration types in our training set. Considering that indeed no such structures were included in our training set and half of the interfaces were created with another potential, we believe that this is a reassuring result.

\section{Genetic algorithm settings}\label{subsec:GA_settings}
Settings for the \gls{GA}~\cite{Weal2021} used to find low-energy \glspl{NP} in Sec.~V.

\begin{widetext}\begin{lstlisting}
# general settings
pop_size = 100
generations = 2500
no_offspring_per_generation = 20

# offspring creation using the mating and mutation procedures
creating_offspring_mode = "Either_Mating_and_Mutation"
crossover_type = "CAS_weighted"
mutation_types = [['random', 0.5], ['random_50', 0.5]]
chance_of_mutation = 0.35

# epoch criterion
epoch_settings = {'epoch mode': 'same population', 'max repeat': 5}

# new cluster creation
r_ij = 4.9
cell_length = r_ij * (sum([float(noAtoms) for noAtoms in \
              list(cluster_makeup.values())]) ** (1.0/3.0))
vacuum_to_add_length = 5.0

# predation scheme
predation_information = {'Predation Operator': 'Energy',
                         'mode': 'comprehensive',
                         'minimum_energy_diff': 0.1 / 1000,
                         'type_of_comprehensive_scheme': 'fitness'}
# fitness scheme
energy_fitness_function = {'function': 'exponential', 'alpha': 3.0}
fitness_information = {'Fitness Operator': 'Energy',
                       'fitness_function': energy_fitness_function}
\end{lstlisting}\end{widetext}

\section{Finnis-Sinclair}
\begin{figure}
\centering
    \includegraphics[width=\linewidth,keepaspectratio]{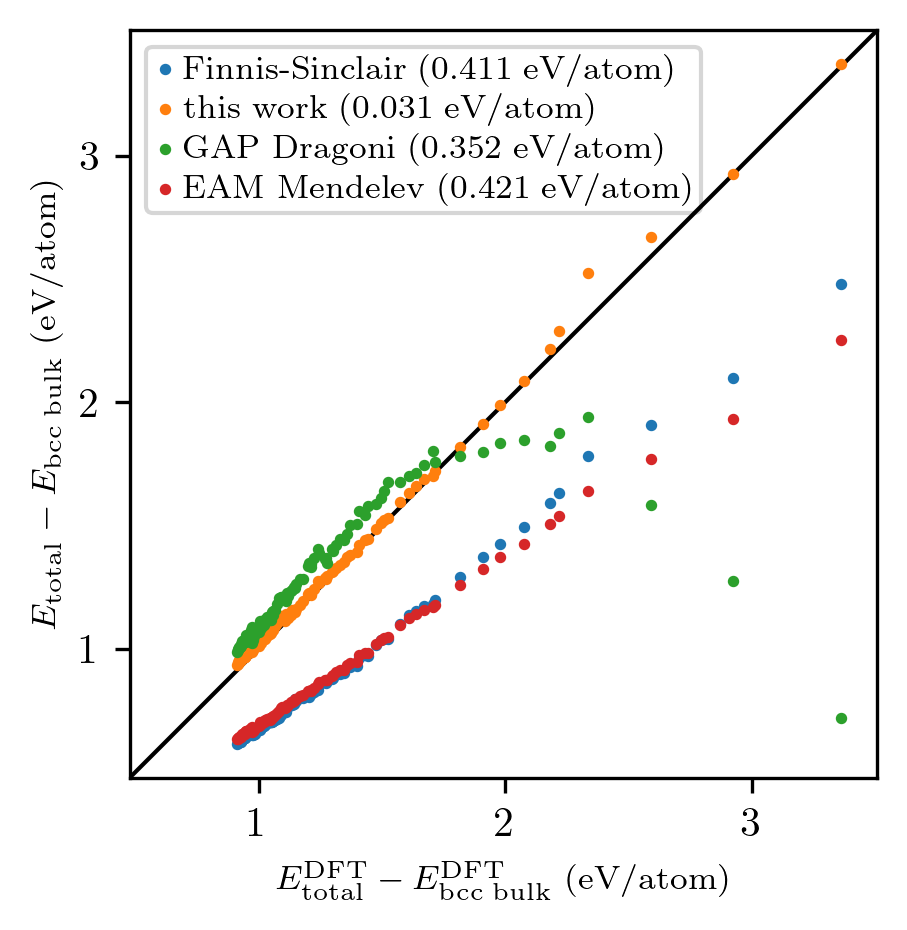}
    \caption{Energies for the \glspl{NP} from the Cambridge Energy Landscape Database: computed with our \gls{GAP}, the Finnis-Sinclair \gls{EAM}, Dragoni \gls{GAP} and Mendelev \gls{EAM} compared to \gls{DFT}.}
    \label{fig_s:Cambridge_FS_vs_GAP}
\end{figure}

To showcase the capabilities of the Finnis-Sinclair \gls{EAM} for \glspl{NP}, we computed the energies of the \glspl{NP} in the \gls{CELD} using \gls{DFT} and the three potentials used in this work.
\fig{fig_s:Cambridge_FS_vs_GAP} shows the errors of the interatomic potentials compared to \gls{DFT}. Only our \gls{GAP} yields acceptable accuracy over the whole size range of \glspl{NP}.
The Dragoni \gls{GAP} shows good accuracy for the largest \glspl{NP}, but fully diverges towards smaller \glspl{NP}. This is likely connected to the spurious surface energies the Dragoni \gls{GAP} yields for \gls{hcp} surfaces.
The Mendelev and Finnis-Sinclair \glspl{EAM} on the other hand show an almost constant offset from the \gls{DFT} values, only slightly increasing towards smaller \glspl{NP}.

\bibliography{references}